\documentclass[journal]{IEEEtran}
\usepackage{graphics,epsf,psfrag,cite,times,wrapfig}
\usepackage{graphicx}
\usepackage{subfigure}
\usepackage{url}
\usepackage{changebar}
\usepackage{fancybox}
\usepackage{graphics}
\usepackage{epsfig,subfigure}
\usepackage{amsmath}
\usepackage{amssymb}
\usepackage{epsfig}
\usepackage{epic}
\usepackage{amsthm}
\usepackage{color}
\usepackage{mathtools}

\makeatletter
\def\ps@headings{%
\def\@oddhead{\mbox{}\scriptsize\rightmark \hfil \thepage}%
\def\@evenhead{\scriptsize\thepage \hfil\leftmark\mbox{}}%
\def\@oddfoot{}%
\def\@evenfoot{}}
\newtheorem*{rep@theorem}{\rep@title}
\newcommand{\newreptheorem}[2]{%
\newenvironment{rep#1}[1]{%
 \def\rep@title{#2 \ref{##1}}%
 \begin{rep@theorem}}%
 {\end{rep@theorem}}}
\makeatother

\makeatother
\newtheorem{theorem}{Theorem}
\newreptheorem{theorem}{Theorem}
\newtheorem{lemma}{Lemma}
\newreptheorem{lemma}{Lemma}

   \newtheoremstyle{example}{\topsep}{\topsep}%
     {}
     {}
     {\bfseries}
     {}
     {\newline}
     {\thmname{#1}\thmnumber{ #2}\thmnote{ #3}}

   \theoremstyle{example}
   \newtheorem{example}{Example}

\renewcommand{\eqref}[1]{(\ref{eq:#1})}

\newcommand{\myitemize}{\begin{itemize}\vspace*{-.05in}}

\newcommand{\myenumerate}{\begin{enumerate}\vspace*{-.1in}}
\newcommand{\ignore}[1]{}

\newcommand{\E}[1]{{\mathbb{E}}\left[{#1}\right]}
\newcommand{\var}[1]{{\mbox{var}}\left({#1}\right)}

\newcommand{\Prob}[1]{{\mathbb{P}}\left({#1}\right)}
\newcommand{\argmin}{\operatornamewithlimits{arg\ min}}
\newcommand{\argmax}{\operatornamewithlimits{arg\ max}}

\newcommand{\comment}[1]{}

\def\BibTeX{{\rmfamily B\kern-.05em{\scshape i\kern-.025em b}\kern-.08em \TeX}}

\def\compactify{\itemsep=0pt \topsep=0pt \partopsep=0pt \parsep=0pt}
\let\latexusecounter=\usecounter

\date{\today}

\title{{Basic Performance Limits and Tradeoffs in Energy Harvesting Sensor Nodes with Finite Data and Energy Storage}}
\author{Rahul~Srivastava,~\IEEEmembership{Member,~IEEE} and~Can~Emre~Koksal,~\IEEEmembership{Member,~IEEE}%
\thanks{Rahul Srivastava is with the Wireless Connectivity Group, Broadcom Corporation, Sunnyvale, CA (e-mail: rahul.srivastava@gmail.com).}
\thanks{Can Emre Koksal is with the Department of Electrical and Computer Engineering,
The Ohio State University, Columbus, OH (e-mail: koksal@ece.osu.edu).}
\thanks{This work was in part supported by NSF Grants CNS 0831919, CCF 0916664, and CNS 1054738.}
}

\begin{document}
\maketitle
\begin{abstract}
As many sensor network applications require deployment in remote and
hard-to-reach areas, it is critical to ensure that such networks are capable
of operating unattended for long durations. Consequently, the concept of using nodes with
energy replenishment capabilities has been gaining popularity.
However, new techniques and protocols must be developed to
maximize the performance of sensor networks with energy replenishment. Here, we analyze limits of the performance of sensor nodes with
limited energy, being replenished at a variable rate. We provide a simple
localized energy management scheme that achieves a performance close to that
with an unlimited energy source, and at the same time keeps the probability of
complete battery discharge low. Based on the insights developed, we address the
problem of energy management for energy-replenishing nodes with finite battery and finite data buffer capacities. To this end, we give an energy management scheme that achieves the optimal utility asymptotically while keeping both the battery discharge and data loss probabilities low.
\end{abstract}

\section{Introduction}
\label{sec:intro}

Advances in wireless networking combined with data acquisition have
enabled us to remotely sense our
environment~\cite{Hartung06MobiSys,Martinez04Secon}. As these
applications may require deployment in hard-to-reach
areas, it is critical to ensure that such networks are capable of
operating with full autonomy for long durations. The lack of a continuous power source in most scenarios and the limited lifetime
of batteries have hindered the deployment of such networks. However, developments in renewable energy sources~\cite{Kansal03ISLPED,Meninger01TVLSI,Paradiso01UbiComp,Rabaey00Computer,Weber03ISLPED,Raghunathan05IPSN}
suggest that it is feasible for sensor networks to operate unattended for extended periods. These renewable sources of energy
typically provide energy replenishment
at a rate that could be variable and dependent on the surroundings.
Examples include, self-powered sensors that rely on
harvesting strain and vibration energies from their working
environment~\cite{Meninger01TVLSI}, as well as sensors with solar
cells~\cite{Paradiso01UbiComp,Rabaey00Computer,Weber03ISLPED}.

In this paper, we analyze the {\em limits} of the {\em performance} of networks
comprised of sensor nodes with limited energy, being replenished at
a variable rate. We provide a simple localized {\em energy management
scheme} that achieves a performance, close to the optimal scheme that has access to an unlimited energy
reservoir. Indeed, we show that, if the performance can be measured by
a general utility function of the energy, under mild assumptions on the replenishment process, it is possible to observe a
polynomial decay for the probability of complete battery discharge, and
at the same time achieve a $\Theta\left(\frac{(\log M)^2}{M^2}\right)$ convergence to the optimal achievable
utility\footnote{The following notations will be used to compare rates of convergence: $a_n=\text{O}(b_n)$ if $a_n$ goes to zero at least as fast as
$b_n$; $a_n=\text{o}(b_n)$ if $a_n$ goes to zero strictly faster than $b_n$; $a_n=\Theta(b_n)$ if $a_n$ and $b_n$ go to zero at the same rate; $a_n=\Omega(b_n)$ if $a_n$ goes to zero no faster than $b_n$.}. Here $M$ is the total capacity of the energy source.
Based on the insights developed, we address the problem of energy management in the presence of a finite data buffer. We modify our basic energy management scheme to achieve a $\Theta\left(\frac{(\log K)^2}{K^2}\right)$ convergence to the maximum utility achievable by a scheme that has access to an infinite data and energy buffers. Here $K$ is the data buffer size. In addition, this scheme achieves an exponential decay with $M$ for the battery discharge probability and a polynomial decay with $K$ for the data loss probability. To evaluate these decay rates, the main tools we use are the \emph{large deviations theory} and \emph{{stochastic process limits}}.

The added dimension of renewable energy makes the problem of energy
management in sensor networks substantially different from its
non-replenishment counterpart.
For nodes with replenishment, conservative energy expenditure may
lead to missed recharging opportunities due to battery capacity
limitations. On the other hand, aggressive usage of energy may cause battery outages that leads to lack of
coverage or connectivity for certain time periods. Thus, new techniques must be
developed to balance these seemingly contradictory goals to maximize performance. Here, our {\em main goal} will be to identify the performance limits of sensor nodes with energy replenishment and provide guidelines to approach these limits.


Many fundamental wireless communication and networking problems can be stated
as {\em utility maximization} problems, subject to energy constraints. The
{utility function} can be the throughput (e.g., in energy efficient
routing), the probability of detection of an intruder (e.g., in coverage) or
the network lifetime (e.g., in sleep-wake scheduling) or the achievable rate
of reliable transmission in basic wireless communication. These problems have
been mainly addressed for stations with unlimited and/or non-replenishing
energy {stores}.
Here, we address the problem of maximizing a utility function of the data transmission rate in the presence of energy replenishment. The solution of the optimization problem requires stochastic optimization techniques involving high computational overheads that might be unsuitable for sensor nodes. Consequently, we will focus our attention on simple localized solutions that achieve near-optimal or asymptotically optimal performance. We use tools from large deviations theory and {stochastic process limits} to find closed-form expressions for the data loss and the battery discharge probabilities. These techniques allow us to analyze our schemes under mild assumptions on the battery charging and data arrival processes.
{There have been recent works that have studied different problems in networks with energy replenishment. Kar, et.~al.,~\cite{Kar06ToN} proposed an activation scheme for rechargeable sensors that maximizes the network-level utility of sensing networks. The utility function in~\cite{Kar06ToN} depends on the number of active sensors. Gatzianas, et.~al.,~~\cite{Gatzianas10TWC} used back pressure policies to maximize the network flow of information in networks with energy replenishment. While~\cite{Kar06ToN,Gatzianas10TWC} look at the total system utility, we will focus on the analyzing node-level performance leading to localized energy management schemes.
Liu, et.~al.,~\cite{Liu10Infocom} derived a battery control scheme similar to the one described in this work. In addition to providing stronger convergence results than the one in~\cite{Liu10Infocom} with sole battery control, we also consider the effect of a finite data buffer in this paper. {Ozel and Ulukus~\cite{Ozel10PIMRC} evaluated the Gaussian channel capacity in the energy harvesting scenario and showed that the capacity is unchanged for a class of replenishment process.}
Kansal, et.~al.,~\cite{Kansal07TECS} introduced the concept of energy neutral operation, wherein the energy consumed by a node is less than or equal to the energy harvested. Vigorito, et.~al.,~\cite{Vigorito07Secon} extended the idea of energy neutral operation to propose an algorithm that attempts to keep the battery state close to a fixed level and at the same time stabilizes the duty cycle in order to maximize system performance. Sharma, et.~al.,~\cite{Sharma10TWC} proposed a throughput optimal energy management scheme for energy harvesting nodes. {Ho and Zhang~\cite{Ho10ISIT} solved the problem of optimal energy allocation in energy harvesting nodes using dynamic programming techniques.} However,~\cite{Kansal07TECS,Vigorito07Secon,Sharma10TWC, Ho10ISIT} do not contain an analytical evaluation of the battery discharge or the data loss probabilities for their energy management schemes.
} 

The outline of this paper is as follows. We first state the general form of the utility maximization problem in
Section~\ref{sec:model} and show ways to achieve the maximum achievable
utility with replenishing sources.
In Section~\ref{sec:joint_queue_and_battery} we add a finite buffer to the problem and study energy management schemes that achieve optimal utility asymptotically while keeping the probabilities of battery discharge and data loss low.
We numerically evaluate the performance of our energy management schemes in Section~\ref{sec:simulations}. We wrap up with conclusions in Section~\ref{sec:conclusion}.

\section{Achieving Maximum Utility With a Finite-Battery Constraint}
\label{sec:model}

\subsection{System Model and Problem Statement}
\noindent
\begin{figure}
\centering
\psfrag{r}[bc]{$r(t)$}
\psfrag{e}[bc]{$e^\mathcal{S}(t)$}
\psfrag{M}[tc]{$M$}
\psfrag{B}[tc]{$B(t)$}
\psfrag{0}[tc]{$0$}
\centerline{\includegraphics[width=0.2\textwidth]{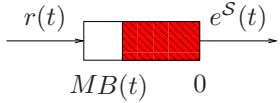}}
\caption{Energy {store} with a replenishment rate $r(t)$.} \label{fig:battery}
\vspace*{-0.2in}
\end{figure}
Fig.~\ref{fig:battery} shows the energy store (or the battery) of a node.
The total capacity of this battery is $M$ units of energy. We denote the total available energy in the battery as $B(t)$, where $t$ is the discrete time index. The battery replenishes at a rate $r(t)$. {The process $\{r(t),t\geq 1\}$ is assumed to be an ergodic stochastic process with a long term mean $\lim_{\tau\to\infty} \frac{1}{\tau}\sum_{t=1}^\tau{r(t)}\overset{\text{a.s.}}\rightarrow\mu$.} An energy management scheme $\mathcal{S}$ draws energy from this battery at a rate $e^\mathcal{S}(t)$ to achieve certain tasks. {The success of the node in achieving these tasks is measured in terms of a utility function $U(e^\mathcal{S}(t))$ of the consumed energy $e^\mathcal{S}(t)$. We assume $U(e)$ to be a concave, non-decreasing\footnote{Note that, in many practical scenarios, it is reasonable to assume that the utility function is non-decreasing and concave, due to diminishing
returns for increasing power.} and analytic function of $e$ over $e\geq 0$.}
We define the time average utility,
\begin{equation}
\label{average_cost}
\bar{U}^\mathcal{S}(\tau)=\frac{1}{\tau} \sum_{t=1}^{\tau} U(e^\mathcal{S}(t)).
\end{equation}

We consider the optimization problem in which a node tries to maximize
its {long-term average utility, $\bar{U}^\mathcal{S}=\liminf_{\tau \to \infty}
\bar{U}^\mathcal{S}(\tau)$}, subject to battery constraints:
\begin{align}
\label{optim_problem_1}
\max_{\left\{e^\mathcal{S}(t),\ t\geq 1\right\}} &\quad \bar{U}^\mathcal{S} \\
\nonumber
{\text{subject to}} &\quad B(t) =
\min \{ M, B(t-1) + r(t)-e^\mathcal{S}(t-1) \}  \\
\nonumber
\text{and} &\quad e^\mathcal{S}(t) \leq B(t).
\end{align}
One approach to solving this optimization problem is by using Markov decision process (MDP) techniques. Since solving MDPs is computationally intensive, these methods may not be suitable for computationally-limited sensor nodes. Consequently, we seek schemes that are easy to implement and yet achieve close to optimal performance.
The next lemma gives an upper bound for the asymptotic time-average utility achieved over all ergodic energy management policies.
\begin{lemma}\label{lemma:utilbound}
Let $\bar{U}^{\mathcal{S}^*}$ be the solution to Problem~(\ref{optim_problem_1}). Then, $\bar{U}^{\mathcal{S}^*} \leq U(\mu)$.
\end{lemma}
The proof of this lemma, given in Appendix~\ref{app:utilboundproof}, uses Jensen's inequality and conservation of energy arguments. Lemma~\ref{lemma:utilbound} tells us that for any ergodic energy management scheme $\mathcal{S}$, $\bar{U}^\mathcal{S}\leq U(\mu)$. With an unlimited energy reservoir (i.e.,~$M=\infty$) and average energy replenishment rate $\mu$, if one uses $e^\mathcal{S}(t)=\mu$ for all $t\geq1$, this upper bound can be achieved.
However, if $M < \infty$,
achieving $\bar{U}^\mathcal{S}=U(\mu)$ using this simple scheme is not possible. Indeed, due to finite energy storage and variability in $r(t)$, $B(t)$ will occasionally get discharged completely. At such instances, $e^\mathcal{S}(t)$ has to be set to $0$, which will reduce the time-average utility.
The question we answer is, ``how close can the average utility $\bar{U}^\mathcal{S}$ get to the upper bound asymptotically, as $M\to\infty$, while keeping the long-term battery discharge rate low?''

\subsection{An Asymptotically Optimal Energy Management Scheme}
\label{sec:near_opt}
In this section we show that there is a trade-off between achieving maximum utility and keeping the discharge rate low.
First, we make some weak assumptions on the replenishment process $r(t)$, which we will be using throughout this paper. In particular, we assume that the asymptotic semi-invariant log moment generating function,
\begin{equation}
\label{log_mmt_gen_fnc}
\bar{\Lambda}_{r}(s)=\lim_{\tau \rightarrow \infty} \frac{1}{\tau}\log \E{\exp\left( s \sum_{t=1}^\tau r(t) \right)},
\end{equation}
of $r(t)$ exists for $s\in (-\infty,s_{\max})$, for some $s_{\max}>0$. We also assume that the asymptotic variance $\bar{\sigma}_r^2\triangleq \lim_{\tau\to\infty}\frac{1}{\tau}\var{\sum_{t=1}^\tau r(t)}$ of $r(t)$ exists\footnote{{Examples of valid processes include the following. 1) Any i.i.d. process with a sample distribution that has finite moments of all orders; 2) All Gaussian processes with an autocovariance function that has a finite integral; 3) The process obtained by adding a deterministic periodic function of time (to mimic the daily cycles of solar radiation) to the aforementioned processes in 1), 2).}}. Note
that, in practice, the recharging process is not necessarily stationary.
While this assumption does allow the possibility that the statistics of $r(t)$ has variations (e.g., due to clouds and the solar power at different times of the day), it rules out the possibility of long-range dependencies in $r(t)$.

From the discussion in previous section, we can infer that by choosing a battery drift, defined as $r(t)-e(t-1)$, that goes to zero with increasing battery size, one might achieve a long-term average utility that is close to $U(\mu)$ as $M$ increases. However, smaller drift away from the empty battery state implies a more frequent occurrence of the complete battery discharge event. In the following theorem, we quantify this tradeoff between the achievable utility and the battery discharge rate, asymptotically in the large battery regime. In this regime, the battery size $M$ is large enough for the variations in $r(t)$ to average out nicely over the time scale that $B(t)$ changes significantly. Consequently, we now define the long-term battery discharge rate as the probability of discharge, i.e., $p_\text{discharge}(M) \triangleq \lim_{\tau \to \infty} \frac{1}{\tau}\sum_{t=1}^\infty {\cal I}_0^B(t)$, where the indicator variable ${\cal I}_0^B(t)=1$ if $B(t)=0$ and is identical to $0$ otherwise.

Next, we show that one can achieve a battery discharge probability that exhibits a polynomial decay {of arbitrary order} with the battery size, and at the same time achieves a
utility that approaches the maximum achievable utility as $(\log M)^2/M^2$.

\begin{theorem}
\label{th:quaddecay}
Consider any continous, concave, non-decreasing, and analytic utility function $U(e(t))$ over the non-negative real line such
that $\left| \frac{\partial^2 U(e)}{\partial e^2}\right| < \infty$ for all $e>0$.
Given any $\beta \geq 2$, there exists an energy management scheme
${\mathcal{B}}$ such that the associated battery discharge probability $p_\text{discharge}^{\mathcal{B}}(M)=\Theta(M^{-\beta})$ and
$U(\mu) - \bar{U}^{\mathcal{B}} =
\Theta \left(\left(\frac{\log M}{M}\right)^2 \right)$.
\end{theorem}

\begin{figure}
\centering
\psfrag{Rate}[bc]{\small{$U(e^\mathcal{S}(t))$}}
\psfrag{power}[cc]{\small{$e^\mathcal{S}(t)$}}
\psfrag{lambda}[cc]{\small{$\mu$}}
\psfrag{u}[cr]{\small{$U^+$}}
\psfrag{d}[cr]{\small{$U^-$}}
\psfrag{delp}[cl]{\small{$\delta^\mathcal{B}$}}
\psfrag{delm}[cr]{\small{$\delta^\mathcal{B}$}}
\centerline{\includegraphics[height=1.35in]{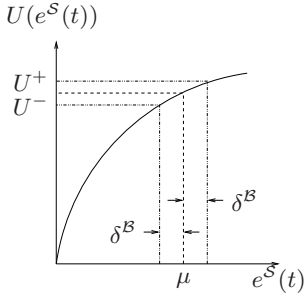}}
\caption{With scheme ${\mathcal{B}}$, utility alternates
between $U^+$ and $U^-$.}
\label{fig:utility_curve}
\vspace*{-0.2in}
\end{figure}
We give a brief sketch of the proof, details of which can be found in
Appendix~\ref{app:prop2}. Our proof is constructive as we show a strategy that achieves the asymptotic convergence rates given in Theorem~\ref{th:quaddecay}. Our scheme is motivated by the buffer control strategy introduced in~\cite{Tse93Allerton} to achieve the near-optimal distortion for variable rate lossy compression. Consider the allocation scheme ${\mathcal{B}}$ in which
{\begin{equation}
\label{allocation}
e^{\mathcal{B}}(t)=\begin{cases} \min\{\mu - \delta^\mathcal{B}, B(t)\} , & B(t) < M/2 \\
                  \mu + \delta^\mathcal{B} , & B(t) \geq M/2 \end{cases} ,
\end{equation}}
for some $\delta^\mathcal{B}>0$. As shown in Fig.~\ref{fig:utility_curve}, the
instantaneous utility associated with Scheme ${\mathcal{B}}$ alternates
between $U^-$ and $U^+$, depending on the battery state.
By choosing $\delta_{1}^\mathcal{B}=\beta\bar\sigma_r^2\frac{\log M}{M}$ for some $\beta\geq2$, we show that long-term maximum utility $U(\mu)$ can be achieved asymptotically while achieving decay, as a polynomial of arbitrarily high order, for the battery discharge probability. We note that while the order of the polynomial decay $\beta$ can be made arbitrarily large, it comes at the expense of slower convergence (by some constant factor) to the maximum utility.

Here, we illustrated that with a simple scheme, it is
possible to achieve desirable scaling laws for the performance of a given task, under the assumption that the asymptotic moment generating function of the replenishment process exists.
To illustrate the theorem we consider a specific example.

\begin{example}
\label{ex:communication}


{\bf Achievable Rate in a Gaussian Channel:} We study the basic limits of point
to point communication with finite but replenishing energy {stores}.
For simplicity, we consider the static Gaussian channel. At time $t$, the transmitter transmits a complex valued
block (vector of symbols) ${\mathbf X}(t)$ of unit power and the receiver receives ${\mathbf Y}(t)$. We have,
\begin{equation}
\label{basic_channel}
{\mathbf Y}(t)=h {\mathbf X}(t) + {\mathbf W}(t) ,
\end{equation}
where the channel gain $h$ is a complex constant and ${\mathbf W}(t)$ is additive Gaussian noise with sample variance $N_0$.
We define the {channel SNR} as $\gamma \triangleq |h|^2 / N_0$. The maximum amount of data that could be reliably communicated~\cite{CoverBook} over this channel with an amount of energy $e(t)$ at time $t$ is:
\begin{equation}
\label{channel_capacity}
C(e(t)) = \log_2 \left(1 + e(t)\gamma \right)\ \text{bits/channel use},
\end{equation}
assuming the block size is long enough so that sufficient averaging of additive
noise is possible. Thus, the rate at which reliable communication can be
achieved {at a given block} is a concave non-decreasing function of the transmit power and it can be viewed as our utility function. Consequently, using a constant power $\mu$, the maximum utility of $\bar{C}=C(\mu)$ can be achieved, which is the famous Gaussian channel capacity result. {Clearly, the capacity is possibly achievable, only if the energy store is infinite.}

{With an energy store that is not capable of providing power at a constant rate (e.g., an energy replenishing battery), one may observe outages due to occurences of complete discharge at times. Thus, for such stores, it is not possible to achieve the aforementioned Gaussian channel capacity. However, we can show that, using our simple energy management scheme, one can achieve an average rate that converges to the capacity at an outage probability that converges to zero asymptotically as $M\to\infty$. We assume that each time slot is large enough for sufficiently long code blocks to be formed.

We simply substitute $U(\cdot)$ with $C(\cdot)$ in Eq.~(\ref{average_cost}) to get the relevant optimization problem. With an unlimited energy store ($M = \infty$) of limited average power $\mu$, the maximum achievable long term average rate is identical to the channel capacity, i.e., $C(\mu) = \log_2(1 + \mu\gamma)$ bits/channel use. By using the energy management scheme ${\cal B}$ given in Eq.~(\ref{allocation}), an average rate $\bar{C}^{\mathcal{B}}$ can be achieved such that $C(\mu) - \bar{C}^{\mathcal{B}} = \Theta \left(\frac{(\log M)^2}{M^2} \right)$ while the battery discharge (i.e., the outage) probability follows $p_\text{discharge}^\mathcal{B}(M)=\Theta(M^{-\beta})$ for any given $\beta\geq 2$.}
\end{example}

\subsection{Basic Limits of Energy Management Schemes}\label{subsec:limitations}
To understand the strength of Theorem~\ref{th:quaddecay}, we note that it is not trivial to achieve decaying discharge probability and maximum utility with increasing battery size. In fact, an ergodic\footnote{An ergodic energy management scheme $e^{\mathcal S}(t)$ is the one that satisfies $\lim_{\tau \to \infty} \frac{1}{\tau}\sum_{t=1}^{\tau}e^{\mathcal S}(t)=\E{e^{\mathcal S}(t)}$} energy management scheme cannot achieve exponential decay in discharge probability and convergence (even asymptotically) to the maximum average utility function simultaneously. We formalize this statement in the following theorem.


\begin{theorem}\label{th:expdecay}
Consider any continous, concave and non-decreasing utility function $U(\cdot)$. If an ergodic energy management scheme $\mathcal{S}$ has a discharge probability $p_\text{discharge}^\mathcal{S}(M)=\Theta(\exp(-\alpha_{c} M))$ for some constant $\alpha_{c}>0$, then the time average utility, $\bar{U}^\mathcal{S}$, for Scheme $\mathcal{S}$ satisfies $U(\mu)-\bar{U}^\mathcal{S}=\Omega(1)$.
\end{theorem}

The proof of this theorem is provided in Appendix~\ref{app:prop1} and it is similar to that of Theorem~\ref{th:quaddecay}. We apply \emph{large deviations} technique to the net drift of the battery process to find the decay rate of $p_\text{discharge}^\mathcal{S}(M)$ with $M$. Jensen's inequality is then used to lower bound the difference between $U(\mu)$ and $\bar{U}^\mathcal{S}$.

So far, we have shown how to maximize a concave non-decreasing
utility function subject to battery constraints. At every point in
time, one should choose a power level as close to the replenishment
rate as the battery constraints allow and this way one can
asymptotically achieve a performance very close to that with
unlimited energy {stores}.
The main limitation of this approach is
that it may not be feasible for some applications in practice.
For instance in many sensor network applications, data is stored in finite buffers for transmission.
Since scheme ${\mathcal{B}}$ does not adapt to the buffer state, this may lead to data losses. To overcome these limitations, in the next section, we investigate energy management schemes with buffer and battery constraints.


\section{Achieving Maximum Utility With Finite Buffer and Battery Constraints}
\label{sec:joint_queue_and_battery}
\subsection{System Model and Problem Statement}
In this section, we extend the problem introduced in Section~\ref{sec:model} to the case when data packets arrive at a node and are kept in a finite buffer before transmission. Hence, the task is to transmit packets arriving at the data buffer without dropping them due to exceeding the buffer capacity. We define $Q(t)$ as the data queue state at time $t$, and the data buffer size is $K<\infty$. The data arrival process $a(t)$, represents the amount of data (in bits) arriving at the data buffer in the time slot $t$. The process $\{a(t),t\geq1\}$ is an ergodic process independent of the energy replenishment process $\{r(t),t\geq1\}$ and $\E{a(\tau)}=\lambda$. We assume that the process $a(t)$ has a finite asymptotic variance $\bar{\sigma}_a^2=\lim_{\tau\to\infty}\frac{1}{\tau}\var{\sum_{t=1}^\tau a(t)}$. The energy replenishment model is the same as used previously. We use $C(\cdot)$ as given in Eq.~(\ref{channel_capacity}) as the rate-power function {(continuous, concave, non-decreasing, and analytic)} for the wireless channel and assume that data is served at that rate as a function of the consumed energy $e(t)$ at time $t$. We also assume that $\lambda<C(\mu)$. {Without this condition, there exists no joint energy and data buffer control policy that can simultaneously keep the long-term battery discharge and data loss rates arbitrarily low asymptotically, as $K, M \to\infty$}.

The objective of an efficient energy management scheme in this case is to maximize the average utility function of the data transmitted subject to battery and data buffer constraints:
{\begin{align}
\label{optim_joint_1}
\max_{e(t),\ t\geq 1} &\quad\liminf_{\tau \to \infty}\frac{1}{\tau} \sum_{t=1}^\tau U_D(C(e(t))) \\
\nonumber
{\text{subject to}} &\quad B(t) =
 \min \{ M, B(t-1) 
+ r(t)-e(t-1) \}  , \\
\nonumber
&\quad Q(t) =
 \min \{ K, Q(t-1) 
+ a(t)- C(e(t-1))  \} , \\
\nonumber
&\quad 0\leq e(t) \leq B(t) \quad \text{and}\quad C(e(t)) \leq
Q(t).
\end{align}}
Here $U_D(C(e))$ is a non-decreasing, concave, and analytic utility gained by transmitting $C(e)$ bits. Since $\lambda<C(\mu)$, we know that $U_D(\lambda)$ is an upper bound on the achievable long-term utility with any energy management scheme. This statement can be proved using Jensen's inequality, following identical steps as the proof of Lemma~\ref{lemma:utilbound} and we skip it to avoid repetition.
\subsection{An Asymptotically Optimal Energy Management Scheme}\label{subsec:asyoptimal}
\noindent
\begin{figure}
\centerline{\includegraphics[height=1.3in]{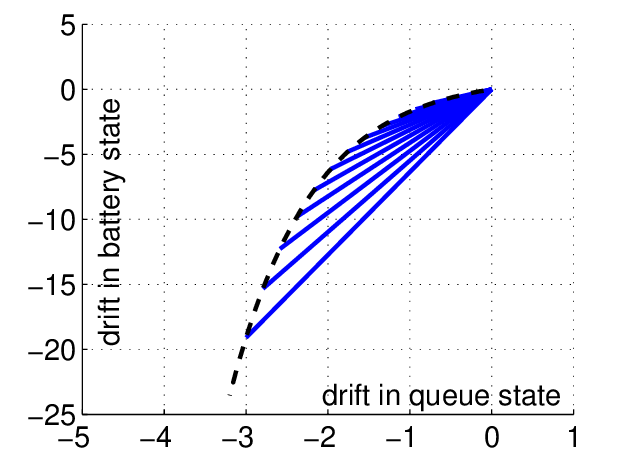}}
\caption{Possible drift directions for $(Q(t),B(t))$ for a Gaussian channel
of channel SNR $0$ dB. Here, at time $t$, $r(t)=0,\ a(t)=0$.}
\label{fig:drifts}
\vspace*{-0.2in}
\end{figure}
Solution of Problem~(\ref{optim_joint_1}) jointly controls the data queue state and the battery state to avoid energy outage and data overflow while maximizing the utility. The main complexity in such an approach stems from
the fact that the drifts of $Q(t)$ and $B(t)$ are dependent.
In Fig.~\ref{fig:drifts} we illustrate the connection between the service rate
and the energy consumed at a time slot $t$ for an Gaussian channel with SNR $0$ dB and $a(t)=0,\ r(t)=0$.
For instance, to provide 3 units of service, the node needs to consume
$\sim 18$ units of energy.

With this dependence, a critical factor one needs to take into consideration is the
relative ``size'' of the data buffer with respect to the battery. In the sequel, we assume a {\em large battery regime},
which implies that,
within the duration that some change occurs in $B(t)$, $Q(t)$ may
fluctuate significantly. Technically, for an Gaussian channel with an
SNR $\gamma$, this assumption implies $M \gg
\frac{1}{\gamma}(2^{\lambda}-1)K$, i.e., the total amount of energy
in the battery is much larger than that required to serve a full
data buffer worth of packets. {In the subsequent asymptotic results, in which both $K,M\to \infty$, the large battery regime implies the following. For all sequences of values, $K_n,M_n$, where both sequence goes to $\infty$ as $n\to\infty$, we assume $K_n/M_n\to 0$ as $n\to\infty$.}

Intuitively, in large battery regime, an energy control algorithm should give
``priority'' to adjusting the queue state to achieve a high performance. Consequently, it should choose $e(t)$ such that the drift of $Q(t)$ is always toward a desired queue state even though this may cause battery drift to be negative. Since battery size is large, such temporary negative drifts are expected to affect the battery discharge rate only minimally.
With these observations, we state the following theorem, which indeed verifies our intuition. This theorem shows an asymptotic tradeoff between the achieved utility and the long-term rates of discharge and data loss as $K\to \infty$. In this regime, the data buffer size is large enough for the variations in $a(t)$ to average out over the time scale that $Q(t)$ changes significantly. {Consequently, we now define the long-term data loss rate as the data loss probability, i.e., $p_\text{loss}^\mathcal{Q}(K) \triangleq \lim_{\tau \to \infty} \frac{1}{\tau}\sum_{t=1}^{\tau} {\cal I}_K^Q(t)$, where the indicator variable ${\cal I}_K^Q(t)=1$ if $Q(t)=K$ and is identical to $0$ otherwise.}

{\begin{theorem}\label{th:bufferbattery}
Consider any non-decreasing concave utility function $U_D(\cdot)$ such that $\left|\frac{\partial^2U_D(C(e))}{\partial e^2}\right|<\infty$ for all $e>0$ {and a rate-power function $C(\cdot)$, both of which are analytic in the non-negative real line}. For any $\lambda < C(\mu)$, {there exists
some $\beta>0$ for which an energy management scheme $\mathcal{Q}$ achieves a data loss probability $p_\text{loss}^\mathcal{Q}(K)=\text{O}(K^{-\beta})$, a battery discharge probability $p_\text{discharge}^\mathcal{Q}(M)=\text{O}(\exp(-\alpha_QM))$ for some $\alpha_Q>0$ and a utility that satisfies $U_D(\lambda)-\bar{U}^\mathcal{Q}=\Theta\left(\frac{(\log{K})^2}{K^2}\right)$ under the large battery regime.}
\end{theorem}

Theorem~\ref{th:bufferbattery} states that it is possible to have an exponential decay (with $M$) for the battery discharge probability and a polynomial decay (with $K$) for the data loss probability and at the same time achieve a time average utility that approaches the upper bound on the achievable long-term utility, $U_D(\lambda)$, as $(\log K)^2/K^2$.} Note that $U_D(\lambda)$ can only be achieved with an infinite battery and data buffer sizes. We provide an outline for the proof, a full version of which can be found in Appendix~\ref{app:BrownianProof}. The proof is constructive as we first present scheme $\mathcal{Q}$, and then derive the performance metrics for this scheme.

\begin{figure}
\centering
\psfrag{Rate}[cc]{\small{Rate}}
\psfrag{power}[cc]{\small{Power}}
\psfrag{delta_a1}[lb]{$\delta^{({a})}$}
\psfrag{delta_a2}[rc]{$\delta^{({a})}$}
\psfrag{lambda_a}[cc]{$\lambda$}
\psfrag{lambda_r}[tc]{$\mu$}
\psfrag{delta_r1}[lc]{$\delta_1^{({r})}$}
\psfrag{delta_r2}[rc]{$\delta_2^{({r})}$}
\psfrag{b1}[cc]{\small{Queue}}
\psfrag{b2}[cc]{$\begin{array}{c}\textrm{\footnotesize{Encoder/}}\\\textrm{\footnotesize{Transmitter}}\end{array}$}
\psfrag{b3}[cc]{\footnotesize{FSMC}}
\psfrag{b4}[cc]{$\begin{array}{c}\textrm{\footnotesize{Decoder/}}\\\textrm{\footnotesize{Receiver}}\end{array}$}
\psfrag{b5}[cc]{$\begin{array}{c}\textrm{\footnotesize{Feedback}}\\\textrm{\footnotesize{Delay $d$}}\end{array}$}
\includegraphics[height=1.35in]{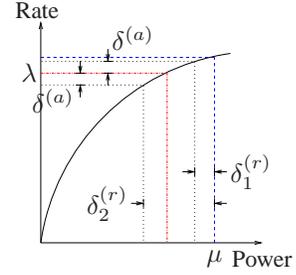}
\vspace{-0.05in}
\caption{Relationship between $\delta^{(a)}$, $\delta_1^{(r)}$ and $\delta_2^{(r)}$.} \label{fig:ratepower}
\vspace*{-0.0in}
\end{figure}
\begin{figure}
\centering
\psfrag{0}[cc]{\small{$0$}}
\psfrag{M}[cr]{\small{$M$}}
\psfrag{M2}[cl]{\small{$({K},{M})$}}
\psfrag{K_2}[cc]{\small{${K}/{2}$}}
\psfrag{K}[cc]{\small{$K$}}
\psfrag{d3}[cl]{$\delta_1^{({r})}$}
\psfrag{d4}[cr]{$\delta_2^{({r})}$}
\psfrag{d1}[bc]{$\delta^{({a})}$}
\psfrag{d2}[bc]{$\delta^{({a})}$}
\psfrag{b1}[cc]{\small{Queue}}
\psfrag{Q}[cc]{$\begin{array}{c}\textrm{\small{Queue}}\\\textrm{\small{State}}\end{array}$}
\psfrag{B}[cc]{$\begin{array}{c}\textrm{\small{Battery}}\\\textrm{\small{State}}\end{array}$}
\psfrag{drifta}[lc]{$\begin{array}{c}\textrm{\small{Queue}}\\\textrm{\small{Drifts}}\end{array}$}
\psfrag{driftb}[lc]{$\begin{array}{c}\textrm{\small{Battery}}\\\textrm{\small{Drifts}}\end{array}$}
\includegraphics[height=1.35in]{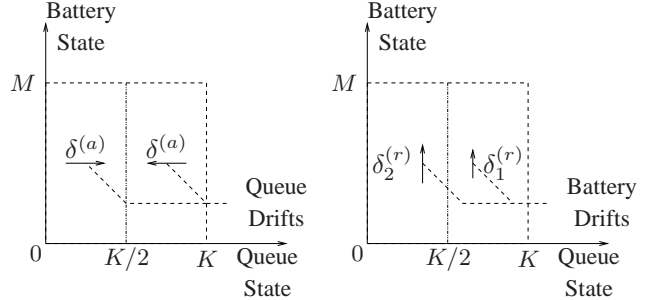}
\vspace{-0.0in}
\caption{{Drifts of the data queue and battery state for scheme~$\mathcal{Q}$.}} \label{fig:drift}
\vspace*{-0.2in}
\end{figure}

Consider the energy management scheme $\mathcal{Q}$, where
\begin{equation}
{
\label{joint_allocation}
e^{\mathcal{Q}}(t) = \begin{cases}
        \min\{\mu - \delta_1^{(r)}, B(t)\}, & Q(t)\geq K/2 \\
        \min\{\mu - \delta_2^{(r)}, B(t)\},  & Q(t) <K/2
\end{cases},
}
\end{equation}
and the drifts $\delta_1^{(r)}$ and $\delta_2^{(r)}$ are chosen to satisfy the relationship
\begin{equation}
\label{eq:delta_choice}
C(\mu-\delta_1^{(r)})-\lambda = \lambda-C(\mu-\delta_2^{(r)}) = \beta_Q\bar\sigma_a^2\frac{\log K}{K},
\end{equation}
{where $\beta_Q$ is constant greater than 2.}
From Fig.~\ref{fig:ratepower}, we note that this choice of energy drifts correspond to a queue drift of $|\delta^{(a)}|=\beta_Q\bar\sigma_a^2\frac{\log K}{K}$, toward the state $K/2$, regardless of the queue state $Q(t)$. The queue and battery drifts with scheme~$\mathcal{Q}$ are illustrated in Fig.~\ref{fig:drift}.
We observe that even though scheme $\mathcal{Q}$ regulates the data queue to a desired state (i.e.,~$K/2$), the battery is always regulated towards full state (i.e.,~$M$). State equation for $Q(t)$ is given by,
\begin{align}
&Q(t+1)\notag\\
&=\begin{cases}\min\{K,Q(t)+a(t)-\lambda-\delta^{({a})}\}, & Q(t)\geq K/2\\
\max\{0,Q(t)+a(t)-\lambda+\delta^{({a})}\}, & Q(t)< K/2\end{cases},\label{eq:emsqueue}
\end{align}
and the state equation for $B(t)$ is given by,
\begin{align}
&B(t+1)\notag\\
&=\begin{cases}\{\min\{M,B(t)+r(t)-\mu+\delta_1^{({r})}\}\}^+, & Q(t)\geq K/2\\
\{\min\{M,B(t)+r(t)-\mu+\delta_2^{({r})}\}\}^+, & Q(t)< K/2\end{cases},\label{eq:emsbattery}
\end{align}
where $\{a\}^+=\max\{0,a\}$.

The main challenge in the proof of Theorem~\ref{th:bufferbattery} is the coupling of the two queues. More specifically, the battery drift in a particular time slot depends on the data queue state, which eliminates the possibility of the application of large deviation techniques for calculating $p_\text{discharge}^\mathcal{Q}(M)$ and $p_\text{loss}^\mathcal{Q}(K)$ difficult. Indeed, closed-form analysis of the stationary distribution for the state of the two-dimensional finite queueing processes is not possible except in some special cases given in~\cite{Harrison87AP}. Since our model does not fall in that category, the steady state probabilities of loss and discharge cannot be derived in closed form. To show the desired order results for Scheme ${\cal Q}$, we transform the problem in two steps as follows:

{\noindent {\bf (T1)} We remove the upper and lower boundaries for the data buffer and the battery respectively, and allow $\{{Q}(t),\ t\geq 0\}$ to take on values in the entire $[0,\infty)$ region and $\{{B}(t),\ t\geq 0\}$ to take on values in the entire $(-\infty,M]$ region. Then, under scheme ${\cal Q}$ as given in (\ref{joint_allocation}), we define $p_\text{overflow}^{\cal Q}(K) \triangleq \lim_{\tau\to\infty} P({Q}(\tau)>K)$ and $p_\text{underflow}^{\cal Q}(M) \triangleq \lim_{\tau\to\infty} P({B}(\tau)<0)$. Using Theorems 1 and 2 in~\cite{Kim01AAP}, one can see that {$p_\text{loss}^{Q}(K)=\text{O}(K^{-\beta})$ for some $\beta>0$} if and only if there exists some {$\beta_Q>0$ such that $p_\text{overflow}^{Q}({K})=\text{O}(K^{-\beta_Q})$}. 
Similarly, one can obtain from Theorem 1 and 2 in~\cite{Kim01AAP} that $p_\text{discharge}^{Q}(M)=\text{O}(\exp(-\alpha_QM))$ if and only if $p_\text{underflow}^{Q}(M)=\text{O}(\exp(-\alpha_QM))$. Thus, it suffices to show the desired scaling laws for the aforementioned unbounded queue state and battery state processes.

\noindent {\bf (T2)} Next, we construct a sequence of arrival rates such that $\lambda\uparrow C(\mu)$. In this limiting regime, from Eq.~(\ref{eq:delta_choice}), $\delta_1^{(r)},\delta_2^{(r)} \downarrow 0$ and hence $\delta^{(a)}\downarrow 0$, for which we also use a sequence of $K$ values that increase to $\infty$ to satisfy the second equality in Eq.~(\ref{eq:delta_choice}). As a result, both the battery and the data queue will operate in the \emph{heavy traffic} limit. We denote the data queue state and the battery state processes in the associated {\em diffusion limit} with $\mathbf{Q}(t)$ and $\mathbf{B}(t)$, respectively. Note that, the probabilities for overshooting the boundaries calculated for the associated diffusion limits, $\Prob{\mathbf{Q}(t)\geq K}$ and $\Prob{\mathbf{B}(t)\leq 0}$ are identical to $\Prob{Q(t)\geq K}$ and $\Prob{B(t)\leq 0}$, respectively in the heavy traffic limit (Chapter 5~\cite{WhittLimits}). Furthermore, since the heavy traffic limit poses a worst case for the probabilities under consideration, the order results of the form $\text{O}(\cdot)$ shown in the heavy traffic limit hold for all $\lambda < C(\mu)$.

However, it is still not straightforward to calculate the associated probabilities in the diffusion limit, since neither $\mathbf{Q}(t)$, nor $\mathbf{B}(t)$ will yield a \emph{Brownian motion} (BM), due to state-dependent variable drifts. To that end, we define upper-half queue state process, $Q_u(t)$, as the queue state process when the state is above $K/2$. This process is formed by taking the sample path of $Q(t)$ and putting the segments for which $Q(t)>K/2$ in a sequence, next to each other (as will be illustrated in Fig.~\ref{fig:qvsqu}). Thus, $Q_u(t) \geq K/2$ for all $t$ with probability $1$. Now, one can see that $\lim_{\tau \to \infty} \Prob{Q_u(\tau)\geq K} \geq p_\text{overflow}^{\cal Q}(K)$. Thus if we prove that $\lim_{\tau \to \infty} \Prob{Q_u(\tau)\geq K} = \text{O}(K^{-\beta_Q})$ for some $\beta_Q > 0$, then it is also true that $p_\text{overflow}^{\cal Q}(K) = \text{O}(K^{-\beta_Q})$ for that $\beta_Q$. The good news is that, since $Q_u(t)$ has a constant drift, under the diffusion limit, it will have a Brownian analogue $\mathbf{Q}_u(t)$, which means that the calculation of the desired probability is easy. Using the properties of BM, we show in Appendix~\ref{app:BrownianProof} that, indeed with Scheme ${\cal Q}$, 
$\lim_{\tau \to \infty} \Prob{\mathbf{Q}_u(\tau)\geq K}= \text{O}(K^{-\beta_Q})$.
To achieve that, we first define a unit reward every time $\mathbf{Q}_u(t)$ goes above $K$. Using renewal-reward theory, we find
\begin{align}
p_\text{overflow}^\mathcal{Q}(K)&=\frac{{\delta^{(a)}}^2}{\bar\sigma_a^2}\exp\left(-\frac{\delta^{({a})}K}{\bar\sigma_a^2}\right).
\end{align}
By substituting $\delta^{(a)}=\beta_Q \bar\sigma_a^2\frac{\log{K}}{K}$, we have the desired scaling law for the queue overflow probability.

Similarly, we denote the diffusion limit of the battery process $B(t)$ by $\mathbf{B}(t)$, which does not constitute a Brownian motion, due to its state-dependent drift. To show the exponential decay rate for the undershoot probability for $\mathbf{B}(t)$, define a BM that lower bounds any given sample path of $\mathbf{B}(t)$. We show that 
$\lim_{\tau \to \infty} \Prob{\mathbf{B}_l(\tau) \leq 0}$ scales as $\text{O}(\exp(-\alpha_QM))$, which implies the same scaling law for the underflow probability of $B(t)$.
Finally, proof for the convergence of the time average utility follows the same line of argument to that for Theorem~\ref{th:quaddecay}.}

\subsection{Exploring Tradeoffs Between Battery Discharge and Buffer Overflow Probabilities}
So far, we focused on achieving performance that was close to the optimal while keeping the probabilities of discharge and data loss low. In this section we look at quantifying tradeoff between the probabilities of battery discharge and data loss.

\begin{theorem}\label{th:expexprate}
{For a channel with a rate-power function $C(\cdot)$ that is continous at $\mu$, there exists an energy management scheme $\mathcal{E}$ that simultaneously achieves
\begin{equation}
\label{eq:thm4_1}
\lim_{M\to \infty}\ \lim_{\lambda\uparrow C(\mu)} \frac{1}{M \delta^{(r)}} \log p_\text{underflow}^\mathcal{E}(M) = -\frac{2}{\bar\sigma_r^2},
\end{equation} 
\begin{equation}
\label{eq:thm4_2}
\lim_{K\to \infty}\ \lim_{\lambda\uparrow C(\mu)} \frac{1}{K \delta^{(a)}} \log p_\text{overflow}^\mathcal{E}(K)=-\frac{2}{\bar\sigma_a^2} ,
\end{equation}
{where $\delta^{(r)} = \nu \left( \mu -C^{-1}(\lambda)\right)$, $\delta^{(a)}=C(\mu-\delta^{(r)})-\lambda$ for any $\nu\in(0,1)$.}
}
\end{theorem}
The proof of this theorem (given in Appendix~\ref{app:expexprate}) is constructive. We consider a energy management scheme $\mathcal{E}$, where,
\begin{align}
\label{eq:thm4_scheme}
e^\mathcal{E}(t)&=\min\{\mu-\delta^{(r)}, B(t)\},
\end{align}
for all $t$, {{with $\delta^{(r)} = \nu \left( \mu -C^{-1}(\lambda)\right)$ for some $\nu\in(0,1)$}. The mean drifts for the battery state and the data queue state are given by $\delta^{(r)}$ and $C(\mu-\delta^{(r)})-\lambda$, respectively. {In the limiting regime $\lambda\uparrow C(\mu)$, 
$\delta^{(r)} \downarrow 0$ and hence $C(\mu-\delta^{(r)})-\lambda \downarrow 0$.}}
{As a result, both the battery and the data queue will operate in the {heavy traffic} limit and we can apply the diffusion limits on these processes to get the required probability results.}
\begin{figure}
\centering
\psfrag{Rate}[cc]{\small{Rate}}
\psfrag{power}[cl]{\small{Power}}
\psfrag{delta_a1}[lb]{$\delta^{({a})}$}
\psfrag{delta_a2}[rc]{$\delta^{({a})}$}
\psfrag{lambda_a}[rc]{$\lambda$}
\psfrag{C}[r]{$C(e^\mathcal{E})$}
\psfrag{lambda_r}[cl]{$\mu$}
\psfrag{e}[cr]{$e^\mathcal{E}$}
\psfrag{delta_r1}[lc]{$\delta^{({r})}$}
\psfrag{delta_r2}[rc]{$\delta^{({r})}$}
\includegraphics[height=1.35in]{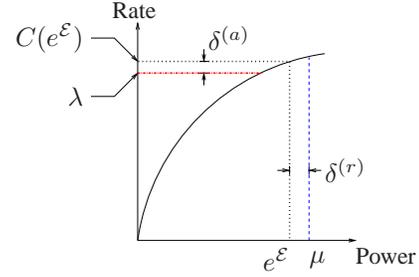}
\caption{Relation between $\delta^{(a)}$ and $\delta^{(r)}$.} \label{fig:ratepower2}
\vspace*{-0.2in}
\end{figure}
Fig.~\ref{fig:ratepower2} illustrates the relationship between $\delta^{(a)}$ and $\delta^{(r)}$.
Any increase in $\delta^{(r)}$ would lead to a corresponding decrease in $\delta^{(a)}$. Since $\delta^{(r)}$ is proportional to the discharge probability decay exponent and $\delta^{(a)}$ is proportional to the data loss probability decay exponent, we will observe the given tradeoff.

Theorem~\ref{th:expexprate} shows that, {in the heavy traffic limit, we observe an exponential decay
for both the battery underflow and the buffer overflow probabilities, with the battery size and the data buffer size respectively.} However, one can also see that, there is a tradeoff in the decay exponents of these two probabilies. More specifically, by varying $\delta^{(r)}$, it is possible to increase (or decrease) the decay exponent for the data loss probability. However this will result a proportional decrease (or increase) in the decay exponent for the battery discharge probability.

\section{Numerical Evaluation}\label{sec:simulations}
Our theorems illustrate tradeoffs for energy management schemes in the buffer and battery size asymptotic regimes and showed optimality of some simple energy management schemes. In this section, we conduct simulations to evaluate the performance of those schemes in the presence of a finite battery and a finite data buffer. We construct the energy replenishment process $r(t)$ using the real solar radiation measurements
collected at the Solar Radiation Research Laboratory~\cite{NRELData}. The data
set used is the global horizontal radiation or the total solar radiation using a Precision Spectral Pyranometer. We use data from January 1999 to July 2010 collected at 1 minute intervals. 

{In our simulations, we chose a battery with storage capacity in the range of $10$-$10^3$ J. {For the replenishment, we considered a 10 cm$^2$ solar panel with $1$\% overall efficiency to get the long-term average of the energy replenishment process $\mu=1.92$ mW. We used the Gaussian channel capacity as the utility function $U(e)=\log(1+\gamma e)$, where channel SNR $\gamma = |h|^2 / N_0$ was defined in Example~\ref{ex:communication}. Here, we take $|h|^2=d^{-\kappa}$, where $d$ is the distance between the transmitter and the receiver and $\kappa$ is the path loss exponent. In the simulations, we let $N_0=-70$ dBm, $d=51.8$ m, and  $\kappa=3.5$, which gives us a mean channel SNR of $12.83$ dB.} Fig.~\ref{fig:AWGN}\subref{fig:subfig1} shows a sample of the solar irradiation process over a 48 hour period.}

\subsection{Battery Constraints with Infinitely Backlogged Buffer}
\begin{figure*}[!htbp]
\centering
\subfigure[Sample of the replenishment process $r(t)$ over a 48 hour period.]{
\includegraphics[width=0.31\textwidth]{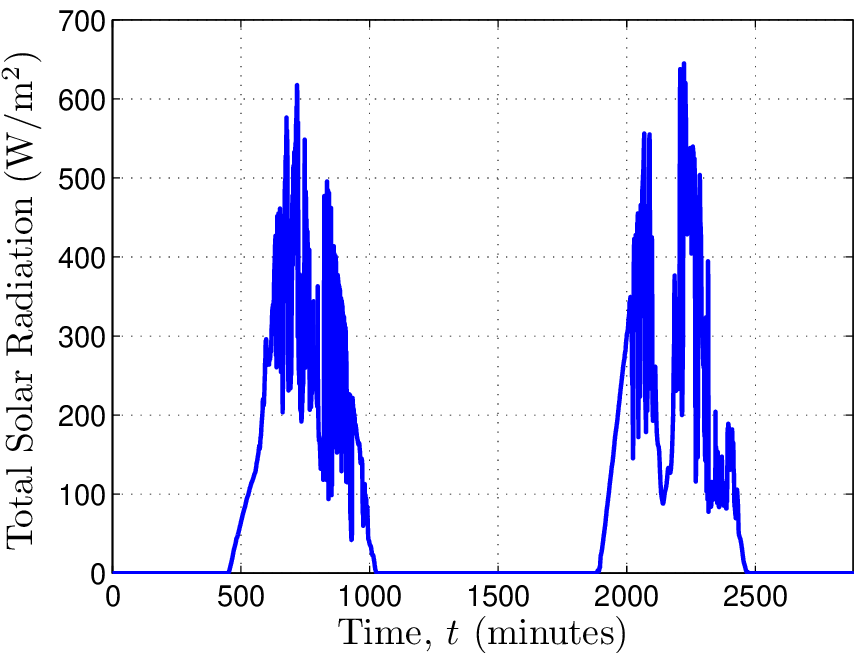}
\label{fig:subfig1}
}
\subfigure[Battery discharge rate scaling with battery size $M$.]{
\includegraphics[width=0.31\textwidth]{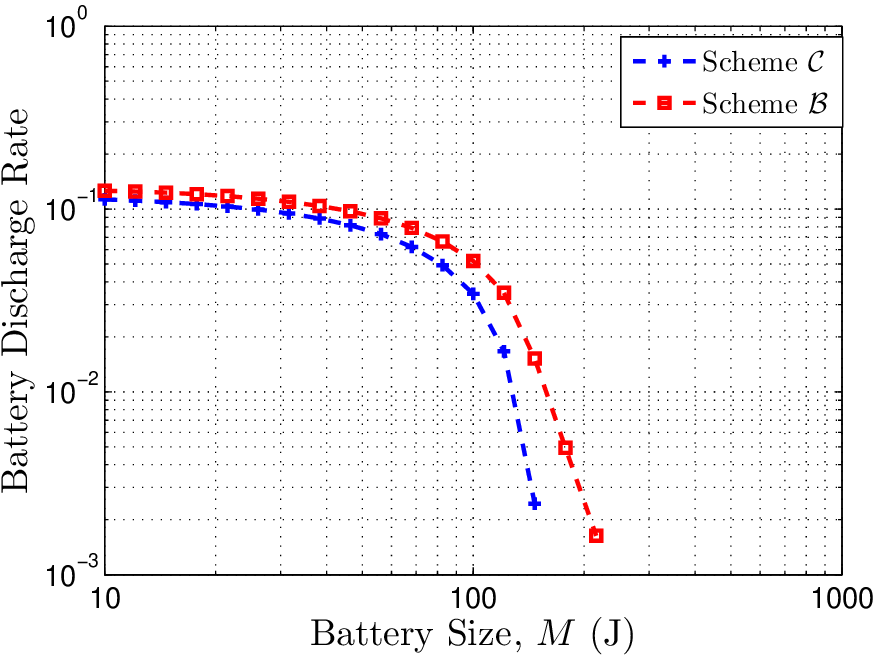}
\label{fig:subfig2}
}
\subfigure[Time average utility scaling with battery size $M$.]{
\includegraphics[width=0.31\textwidth]{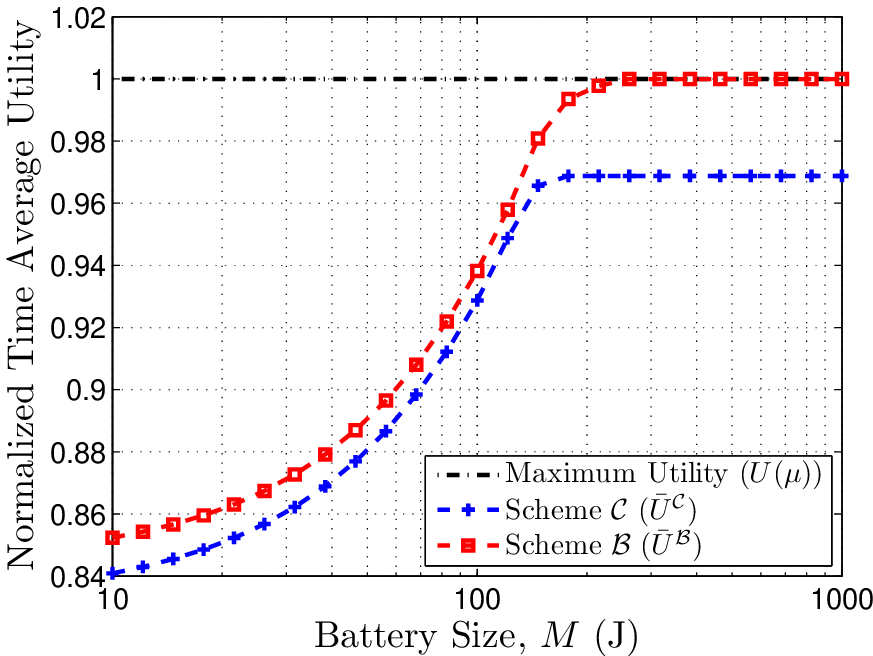}
\label{fig:subfig3}
}
\caption[Optional caption for list of figures]{Performance evaluation for the Gaussian channel example.
}\label{fig:AWGN}
\end{figure*}
{In Fig.~\ref{fig:AWGN}, we revisit the energy management scheme $\mathcal{B}$ discussed in Example~\ref{ex:communication} for an infinitely backlogged data buffer. The communication channel is Gaussian and we choose the polynomial decay exponent $\beta=2$. {To illustrate Theorems~\ref{th:quaddecay}~and~\ref{th:expdecay} simultaneously, we define an energy management scheme $\mathcal{C}$, which allocates a power strictly less than the average replenishment rate. In particular, $e^\mathcal{C}(t)=\{\min\{B(t), \mu-c\}\}^+$, where $c$ is a constant.}
From Theorems~\ref{th:quaddecay} and~\ref{th:expdecay} we know that policy $\mathcal{C}$ achieves an exponential decay for discharge probability compared to the quadratic decay for scheme $\mathcal{B}$. On the other hand, policy $\mathcal{C}$ can not achieve the maximum utility while the utility achieved by scheme $\mathcal{B}$ should {approach} maximum utility as $(\log M)^2/M^2$. 
Fig.~\ref{fig:AWGN}\subref{fig:subfig2} plots the battery discharge rate as a function of the battery size. As expected, policy $\mathcal{C}$ performs better than scheme $\mathcal{B}$. However, the advantage of using policy $\mathcal{B}$ is evident in Fig.~\ref{fig:AWGN}\subref{fig:subfig3}, which compares the normalized time average utilities, $\frac{\bar{U}^{\mathcal{C}}}{U(\mu)}$ and $\frac{\bar{U}^{\mathcal{B}}}{U(\mu)}$, achieved by each scheme. It can be seen that, for the choice of parameters used in this simulation, scheme $\mathcal{B}$ achieves the maximum utility $U(\mu)$ for a battery size of $250$ J, whereas the scheme $\mathcal{C}$ does not achieve the maximum utility even asymptotically.}

\subsection{Buffer and Battery Constraints}
\begin{figure*}[!htbp]
\centering
\subfigure[Battery discharge rate scaling with battery size $M$.]{
\includegraphics[width=0.31\textwidth]{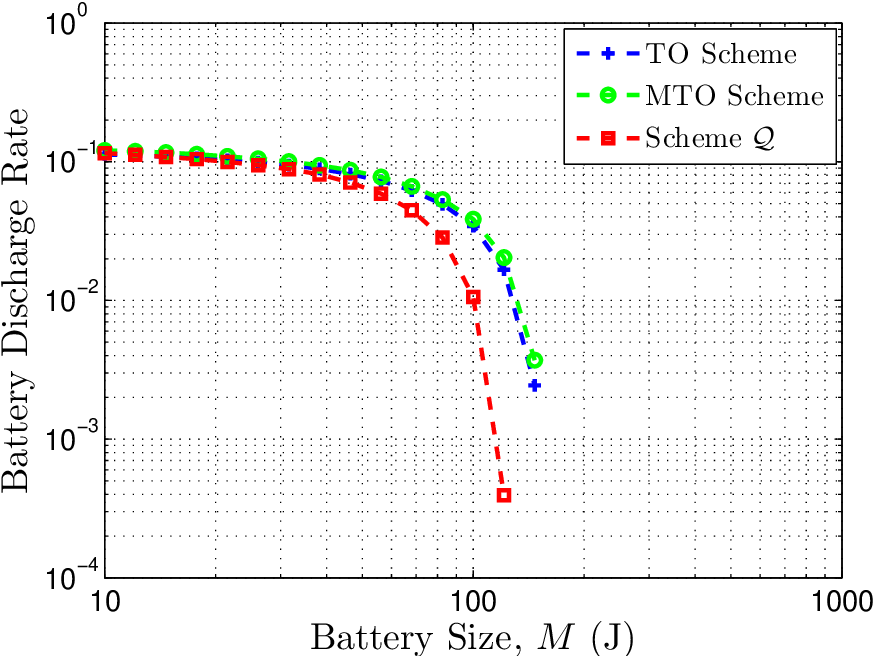}
\label{fig:2subfig1}
}
\subfigure[Data loss rate scaling with buffer size $K$.]{
\includegraphics[width=0.31\textwidth]{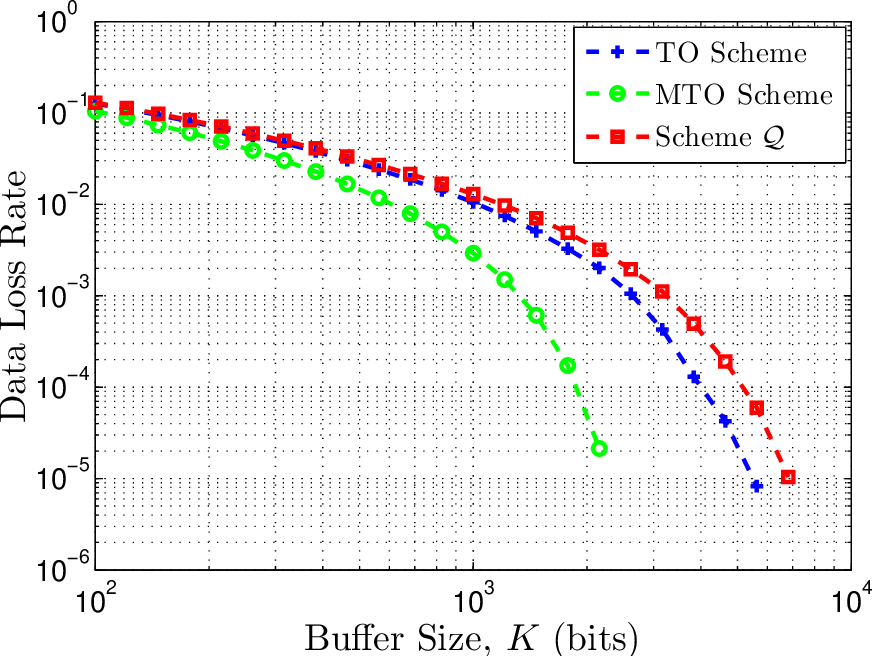}
\label{fig:2subfig2}
}
\subfigure[Time average utility scaling with buffer size $K$.]{
\includegraphics[width=0.31\textwidth]{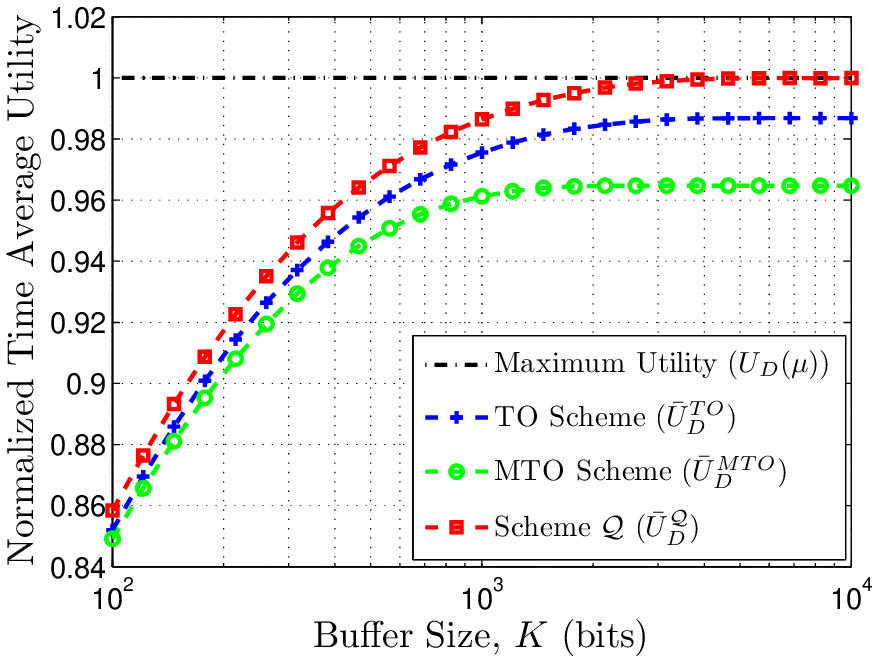}
\label{fig:2subfig3}
}
\caption[Optional caption for list of figures]{Performance evaluation for energy management schemes under buffer and battery constraints.}\label{fig:joint}
\end{figure*}
{Fig.~\ref{fig:joint} compares the performance of energy management schemes when both battery and buffer constraints are present. We simulate the data arrival process by generating a Markov-modulated Poisson process with mean $\lambda=4.12$ bits per time slot. We use a two-state Markov chain to generate a bursty data arrival process. One state of the Markov chain generates a Poisson random variable with mean 10 bits and the other state generates a Poisson random variable with mean 1. {The data utility function is chosen as $U_D(x)=\log_2\left(1+\frac{10x}{\lambda}\right)$. As previously, $\mu=1.92$ mW and we choose $\beta=10$.} 

{To compare with our scheme, we also simulated the performances of Throughput-Optimal (TO) and modified TO (MTO) policies policy given in Eq.~(4) and Eq.~(6) of~\cite{Sharma10TWC} respectively. The energy allocation by TO policy, {which is an instance of policy $\mathcal{C}$}, is specified as:
\begin{align}
e^{TO}(t)&=\min\{B(t),\mu-\epsilon\},\label{eq:TO}
\end{align}
where $\epsilon$ is a constant such that $C(\mu-\epsilon)>\lambda$;} and the energy allocation for MTO policy is given by:
\begin{align}
e^{MTO}(t)=&\min\{C^{-1}(Q(t)),B(t),\notag\\
&\quad\quad0.99(\mu+0.001[B(t)-0.1Q(t))^+]\}.\label{eq:MTO}
\end{align}

In Fig.~\ref{fig:joint}\subref{fig:2subfig1}, we fix the buffer size to $10^4$ bits and plot the battery discharge rate as a function of the battery size $M$. The discharge rates should decay exponentially for both schemes. However, the decay exponent for scheme $\mathcal{Q}$ is larger than the decay exponent for the TO scheme. 
Indeed, Theorem~\ref{th:expexprate} shows that, in the heavy traffic limit, the decay exponent for the discharge probability is proportional to the drift of the battery state. Thus, for small values of $\mu-C^{-1}(\lambda)$, the decay exponent for scheme $\mathcal{Q}$ is approximately proportional to $\mu-C^{-1}(\lambda)$, while the decay exponent for the TO scheme is approximately proportional to $\epsilon < \mu-C^{-1}(\lambda)$, since $\epsilon$ is the drift of the battery state as given in Eq.~(\ref{eq:TO}).

In Fig.~\ref{fig:joint}\subref{fig:2subfig2}, we plot the data loss rate as a function of the buffer size while keeping the battery size fixed at $10^3$ J. We observe that the loss rate for the TO scheme decays faster than that for scheme $\mathcal{Q}$. This trend is expected as the TO scheme should have an exponential decay compared to a quadratic decay for scheme $\mathcal{Q}$. Achieving an exponential decay in the data loss rate comes at the cost of reduced average utility for the TO scheme. On the other hand, the MTO scheme is designed to achieve a low average data queue length. As a result, we observe that the data loss rate of the MTO scheme is lower than the other two schemes. However, the MTO scheme pays the price of slightly higher battery discharge rate.  
Fig.~\ref{fig:joint}\subref{fig:2subfig3} compares the convergence of the time average utilities to the maximum utility function for the two schemes.  
\ignore{Here the utility is defined by a clipper function,
\begin{align*}
U_D(x)=\begin{cases}
x/\lambda & x\leq \lambda\\
1 & x>\lambda
\end{cases}.
\end{align*}}
We observe that scheme $\mathcal{Q}$ converges to the maximum utility for larger buffer sizes ($\sim5000$ bits). On the other hand, TO and MTO schemes do not achieve the optimal utility.}

\begin{figure*}[!htbp]
\centering
\subfigure[Battery discharge rate scaling with traffic intensity $\rho$.]{
\includegraphics[width=0.31\textwidth]{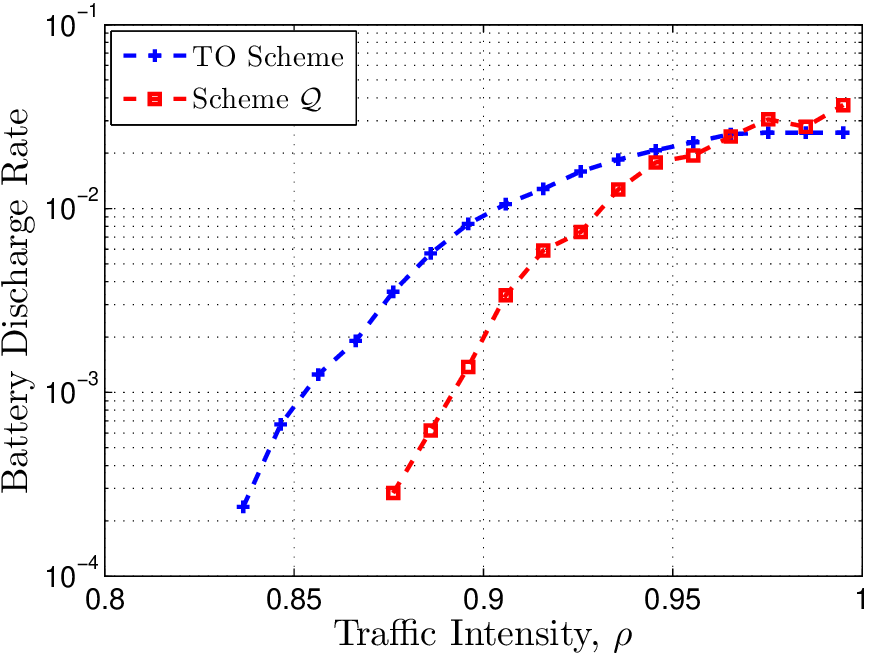}
\label{fig:3subfig1}
}
\subfigure[Data loss rate scaling with traffic intensity $\rho$.]{
\includegraphics[width=0.31\textwidth]{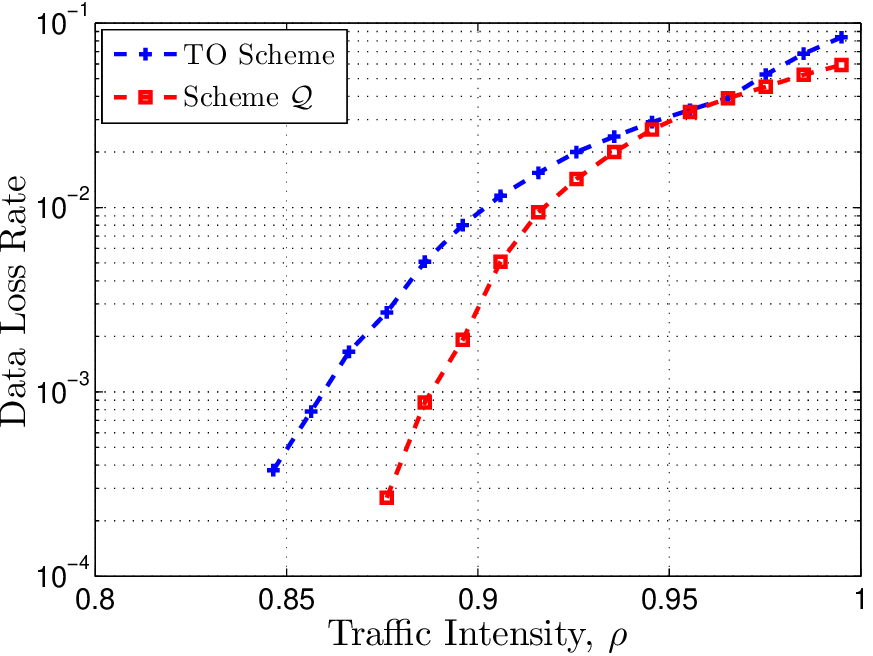}
\label{fig:3subfig2}
}
\subfigure[Time average utility scaling with traffic intensity $\rho$.]{
\includegraphics[width=0.31\textwidth]{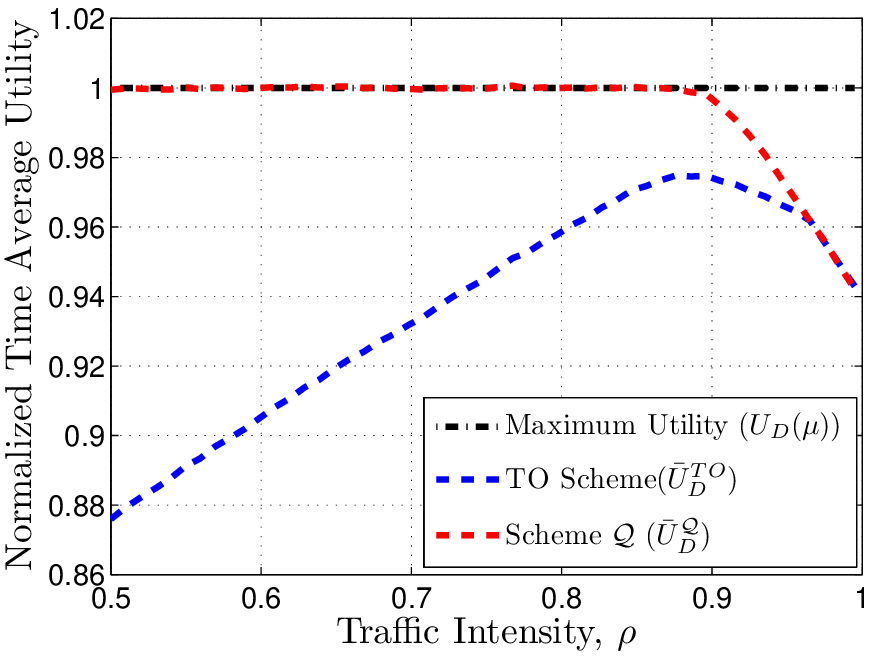}
\label{fig:3subfig3}
}
\caption[Optional caption for list of figures]{Performance evaluation of energy management schemes under buffer and battery constraints with increasing traffic intensities.}\label{fig:intensity}
\end{figure*}
{Fig.~\ref{fig:intensity} compares the performance of energy management schemes with increasing traffic intensity. We define traffic intensity as $\rho\triangleq \frac{\lambda}{C(\mu)}=\frac{\lambda}{\log_2(1+\gamma\mu)}$. We fix the buffer length at $2000$ bits and battery capacity is set at $100$ J.
In Fig.~\ref{fig:intensity}\subref{fig:3subfig1} we observe that the discharge rate increases with traffic intensity. For values of $\rho=0.87$, scheme $\mathcal{Q}$ performs almost an order of magnitude better than the TO scheme in terms of the discharge rate. For traffic intensities close to unity the scheme $\mathcal{Q}$ degenerates to the TO scheme and their performances converge. Fig.~\ref{fig:intensity}\subref{fig:3subfig2} shows that the data loss rates for both schemes also increases with increasing traffic intensity. Similar to the discharge rate, for values of $\rho=0.87$, the loss rate for scheme $\mathcal{Q}$ is almost an order of magnitude lower than that for the TO scheme. This can be explained by the significantly higher discharge rate for the TO scheme leading to severe performance degradation. Finally, we observe in Fig.~\ref{fig:intensity}\subref{fig:3subfig3} that the TO scheme achieves a low average utility at low traffic intensities ($\rho<0.87$). This could be due to the combination of the data buffer getting cleared very often and concavity of the utility function leading to a lower average utility. On the other hand, scheme $\mathcal{Q}$ regulates the buffer level to a non-empty level that ensures that it has data to transmit in most time slots.
As $\rho \to 1$, the performances of the two energy management schemes degrade highly. This is a direct consequence of increasing battery discharge and data loss rates leading to sub-optimal performance.

\subsection{Trade-offs Between Buffer Overflow and Battery Discharge Probabilities}
\begin{figure*}[!htbp]
\centering
\subfigure[Trade-off between battery discharge and data loss probabilities where $E_1=\frac{\bar\sigma_r^2}{2M\mu}$, $E_2=\frac{\bar\sigma_a^2}{2K\lambda}$.]{
\includegraphics[width=0.31\textwidth]{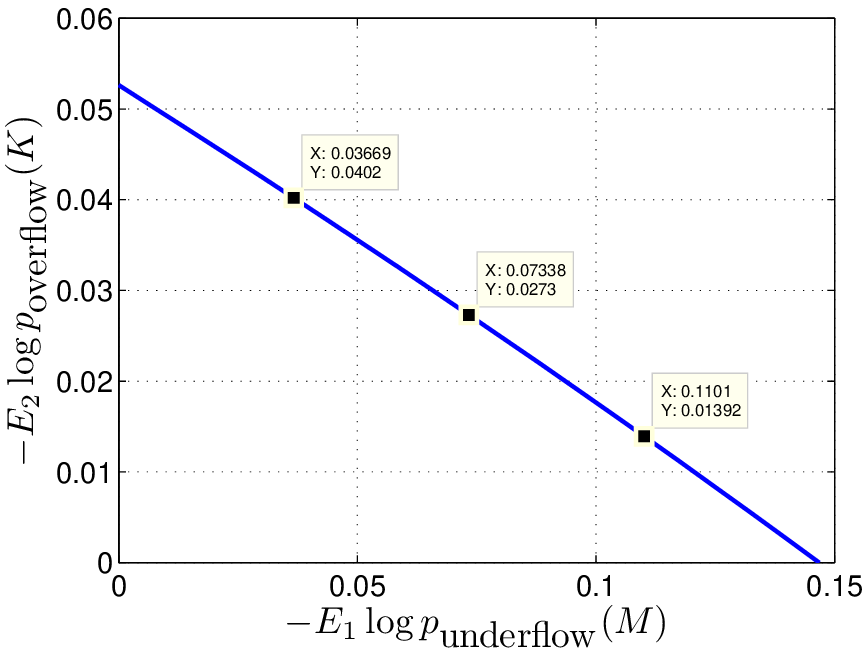}
\label{fig:4subfig1}
}
\subfigure[Data loss rate scaling for different operating points.]{
\includegraphics[width=0.31\textwidth]{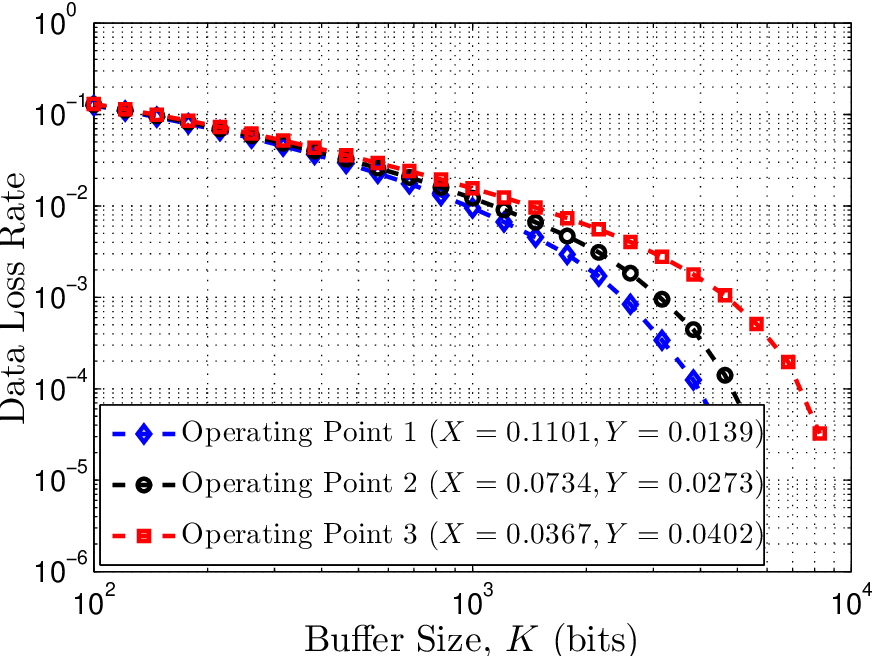}
\label{fig:4subfig2}
}
\subfigure[Battery discharge rate scaling for different operating points.]{
\includegraphics[width=0.31\textwidth]{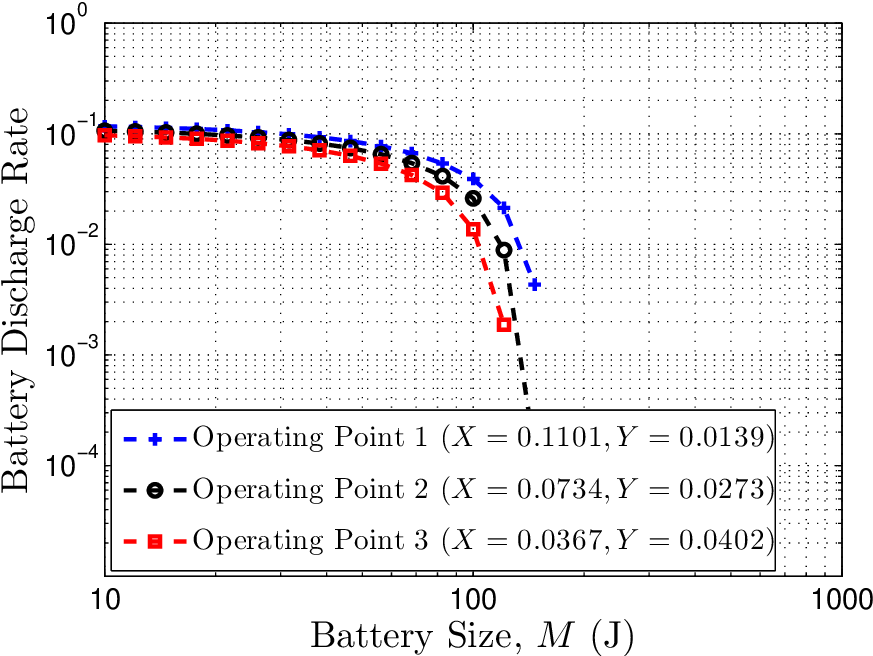}
\label{fig:4subfig3}
}
\caption[Optional caption for list of figures]{Performance evaluation of energy management schemes under buffer and battery constraints with increasing traffic intensities. Here $X=\frac{\bar\sigma_r^2}{2M\mu}(\log_e p_\text{discharge}(M))$ and $Y=\frac{\bar\sigma_a^2}{2K\lambda}(\log_e p_\text{overflow}(K))$.}\label{fig:tradeoff}
\end{figure*}
In Fig.~\ref{fig:tradeoff}, we evaluate the trade-off between {battery overflow and buffer underflow} given in Theorem~\ref{th:expexprate}. We use the data arrival and energy replenishment process used previously. Fig~\ref{fig:4subfig1} illustrates that in order to increase the decay exponent for the {battery underflow probability}, the energy management scheme has to decrease the exponent for the {buffer overflow probability}. We choose three operating points on this curve and evaluate the {battery underflow and buffer overflow} scaling for these points. As we go from operating point 1 to 3, the {buffer underflow decay exponent increases and the battery underflow decay exponent decreases}. In Fig.~\ref{fig:4subfig2}, we observe that the quickest decay for the loss rate is for operating point 1. In Fig.~\ref{fig:4subfig3}, as expected, we see the opposite effect wherein the discharge rate decays fastest for the operating point 3.}

\section{Conclusions}
\label{sec:conclusion}

In this paper, we studied the basic limits and associated tradeoffs for energy management schemes in energy replenishing sensor networks.
{We showed that it is possible to observe a polynomial decay of arbitrary order for the
discharge probability with increased battery size $M$, and at the same time
achieve $\Theta((\log M)^2/M^2)$ convergence to the maximum achievable utility
using a simple energy management scheme.} We showed the strength of this result by showing that it is not possible to simultaneously observe an
exponential decay for the discharge probability and achieve maximum utility.
With the insights drawn, we addressed the problem of energy management with buffer and battery constraints. {We showed that with a finite data buffer of size $K$, in addition to achieving $\Theta((\log K)^2/K^2)$ convergence to the optimum utility, it is possible to achieve a polynomial decay for the data loss probability and exponential decay for the the battery discharge probability using a simple energy management scheme.}

{To analyze the buffer and battery processes we made use of large deviations theory and diffusion approximations. The main advantage of using these tools in our work is that it allows analytical tractability while keeping the system model fairly general in nature.}
Finally, we numerically illustrated the performance of the our simple energy management schemes along with that of another existing scheme, and demonstrated that our scheme can perform up to an order of magnitude better in terms of outage probabilities while achieving the maximum utility asymptotically.

\appendices
\section{Proof of Lemma~\ref{lemma:utilbound}}\label{app:utilboundproof}
To prove this lemma, we first use the finite form of Jensen's inequality to establish,
\begin{align}
\frac{1}{\tau}\sum_{t=1}^\tau U(e^\mathcal{S}(t))\leq U\left(\frac{1}{\tau}\sum_{t=1}^\tau e^\mathcal{S}(t)\right).\notag
\end{align}
Since this inequality holds for any finite $\tau$, passing the limit $\tau\to\infty$, the inequality is preserved,
{\begin{align}
\liminf_{\tau\to\infty}\frac{1}{\tau}\sum_{t=1}^\tau U(e^\mathcal{S}(t))&\leq \liminf_{\tau\to\infty}U\left(\frac{1}{\tau}\sum_{t=1}^\tau e^\mathcal{S}(t)\right)\notag \\
&= U\left(\liminf_{\tau\to\infty}\frac{1}{\tau}\sum_{t=1}^\tau e^\mathcal{S}(t)\right),\label{eq:infjensen}
\end{align}}
where (\ref{eq:infjensen}) follows since $U(\cdot)$ is a continuous function~\cite{RudinAnalysisBook}. From conservation of energy, we have
\begin{align}
\liminf_{\tau\to\infty}\frac{1}{\tau}\sum_{t=1}^\tau e^\mathcal{S}(t)\leq\liminf_{\tau\to\infty}\frac{1}{\tau}\sum_{t=1}^\tau r(t)=\lim_{\tau\to\infty}\frac{1}{\tau}\sum_{t=1}^\tau r(t),\label{eq:energyconserve}
\end{align}
since $M<\infty$. Combining Eqs.~(\ref{eq:infjensen}) and~(\ref{eq:energyconserve}), we have the required result,
\begin{align}
\liminf_{\tau\to\infty}\frac{1}{\tau}\sum_{t=1}^\tau U(e^\mathcal{S}(t))=\bar{U}^\mathcal{S}\leq U(\mu).
\end{align}

\section{Proof of Theorem~\ref{th:quaddecay}}\label{app:prop2}
In this appendix, we prove that the energy management scheme $\mathcal{B}$ achieves the scaling properties given in Theorem~\ref{th:quaddecay}. First, consider a general form of Scheme $\mathcal{B}$:
\begin{align}
e^\mathcal{B}(t)=\begin{cases}
\mu-\delta^-, & B(t)\leq M/2\\
\mu+\delta^+, & B(t)>M/2
\end{cases},
\end{align}
for some pair $\delta^-,\delta^+$, that will be chosen later. We will show that the desired solution involves $\delta^-=\delta^+=\delta^{\cal B}$.

Depending on whether the battery state is less than (or more than) half full, the expected drift of the battery state becomes positive (or negative). Given $B(t)\leq M/2$, the asymptotic semi-invariant log-moment generating function of the battery state drift, $d^-(t)\triangleq r(t)-(\mu-\delta^-)$, is
\begin{align}
\bar\Lambda_{d^-}(s)&=\lim_{\tau\to\infty}\frac{1}{\tau}\log\E{\exp\left(s\sum_{t=1}^\tau d^-(t)\right)}\notag\\
&=\bar\Lambda_r(s)-s(\mu-\delta^-).
\end{align}
Where $\bar\Lambda_r(s)$ is given by Eq.~(\ref{log_mmt_gen_fnc}).
Let $s_{d^-}^*$ be the negative root\footnote{Note that $\bar{\Lambda}_{d^-}(0)=0$ and $\left.\frac{\partial \bar{\Lambda}_{d^-}(s)}{\partial s}\right|_{s=0}=\lim_{T\to\infty}\frac{1}{T}\sum_{t=1}^T\E{d^-(t)}=\delta^->0$. Consequently $s_{d^-}^*<0$ will exist.} of $\bar{\Lambda}_{d^-}(s)$, i.e.,~$\bar\Lambda_{d^-}(s_{d^-}^*)=\bar\Lambda_r(s_{d^-}^*)-s_{d^-}^*(\mu-\delta^-)=0$. Also as $\delta^-\to0$, $s_{d^-}^*\to0$.

Before we prove Theorem~\ref{th:quaddecay}, we state and prove the following lemmas. Lemma~\ref{lemma:exprate2} gives the rate of decay of the probability of battery discharge with respect to the battery size $M$ for the Scheme $\mathcal{B}$. Lemma~\ref{lemma:s} expresses the rate decay exponent $s_{d^-}^*$ for scheme $\mathcal{B}$ in terms of the asymptotic variance of energy replenishment process $r(t)$.

\begin{lemma}\label{lemma:exprate2}
The probability of battery discharge under Scheme $\mathcal{B}$ with battery size $M$ follows $p_\text{discharge}^{\mathcal{B}}(M)=\Theta\left(\exp\left(\frac{s_{d^-}^*M}{2}\right)\right)$, where $s_{d^-}^*$ is the negative root of $\bar{\mu}_{d^-}(s)$.
\end{lemma}
\begin{IEEEproof}
Fix a constant $A>0$ and decompose the time line into intervals, such that each interval is of length $\lceil \frac{M}{2A}\rceil$ and the $i$th interval ends at time slot $t_i=i\lceil\frac{M}{2A}\rceil$. Assume that the system has been active since $t=-\infty$. We define $E_i$ as the event that the battery is empty at the end of time slot 0 and the last time the battery was half full (i.e.,~$M/2$) is some instant during the interval $-i=\left[-(i+1)\lceil\frac{M}{2A}\rceil+1,-i\lceil\frac{M}{2A}\rceil\right]$. The event of an empty battery at time slot 0 can be decomposed as a union of events $E_i$,
\begin{align}
p_\text{discharge}^{\mathcal{B}}(M) &= \sum_{i=0}^\infty \Prob{E_i}\label{eq:probsum2}.
\end{align}

A necessary condition for event $E_i$ to occur is,
\begin{align}
\sum_{t=-(i+1)\lceil\frac{M}{2A}\rceil+1}^0 \left(e^\mathcal{B}(t)-r(t)\right)>\frac{M}{2}.
\end{align}

Using Chernoff's bound, for any $\theta_i\geq0$,
\begin{align}
&\Prob{\sum_{t=-(i+1)\lceil\frac{M}{2A}\rceil+1}^0(e^\mathcal{B}(t)-r(t))>\frac{M}{2}} \notag\\
&\leq\E{\exp\left(\theta_i\sum_{t=-(i+1)\lceil\frac{M}{2A}\rceil+1}^0(e^\mathcal{B}(t)-r(t))\right)}\exp\left(-\theta_i\frac{M}{2}\right) \notag\\
&=\E{\exp\left(-\theta_i\sum_{t=-(i+1)\lceil\frac{M}{2A}\rceil+1}^0 r(t)\right)}\notag\\
&\quad\times\exp\left(\theta_i(i+1)\left\lceil\frac{M}{2A}\right\rceil(\mu-\delta^-)\right)\exp\left(-\theta_i\frac{M}{2}\right)\notag\\
&=\exp\Bigg(-\frac{M}{2}\Bigg[\theta_i\left(1-\frac{i+1}{A}(\mu-\delta^-)\right)-\frac{i+1}{A}\bar\Lambda_{r}(-\theta_i)\notag\\
&\qquad\qquad+\epsilon_i(M,\theta_i)\Bigg]\Bigg),
\end{align}
where $\epsilon_i(M,\theta_i)\rightarrow0$ as $M\rightarrow\infty$.

In order to find the tightest bound for each $i$, we choose $\theta_i^*\geq 0$ to maximize,
\begin{align}
f_i(\theta)\triangleq\theta\left(1-\frac{i+1}{A}(\mu-\delta^-)\right)-\frac{i+1}{A}\bar\Lambda_{r}(-\theta),
\end{align}
over all $\theta>0$ and let $\gamma=\inf_{i\geq0}\sup_{\theta\geq0}f_i(\theta)=\inf_{i\geq0}f_i(\theta_i^*)$.
We can rewrite $f_i(\theta)$ as,
\begin{align}
f_i(\theta)&=\theta-\frac{\mu-\delta^-}{A}\theta-\frac{\bar\Lambda_{r}(-\theta)}{A}-i\left(\frac{(\mu-\delta^-){\theta}+\bar\Lambda_{r}(-{\theta})}{A}\right).\notag
\end{align}
\begin{figure}
\centering
\psfrag{0}[cc]{$0$}
\psfrag{i}[cc]{$i$}
\psfrag{K}[cc]{$J$}
\psfrag{a}[lc]{$\gamma+i\beta$}
\psfrag{c}[cc]{$\gamma$}
\psfrag{b}[lc]{$f_i(\tilde{\theta})$}
\psfrag{0}[cc]{$0$}
\psfrag{slope}[cc]{slope=$\text{E}\left[X_i\right]=-\delta_1^{(a)}$}
\psfrag{min1}[rc]{$\inf_{r\leq r_{\text{buffer}}^*}\mu(r)$}
\psfrag{min2}[lc]{$-\mu_{r_{\text{buffer}}^*}^*(0)$}
\includegraphics[width=0.3\textwidth]{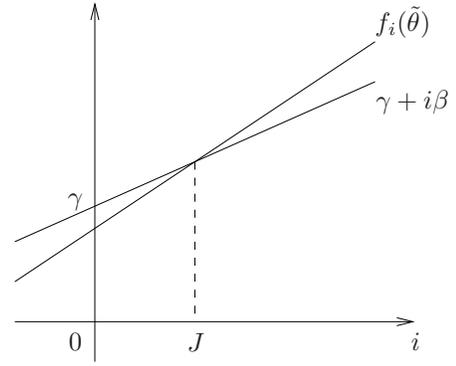}
\caption{A geometric proof for the existence of $J$ and $\delta>0$ such that for every $i>J$, $f_i(\tilde{\theta})>\gamma+i\delta$.} \label{fig:diffslopeproof}
\vspace*{-0.2in}
\end{figure}
Since $\lim_{\tau\to\infty}\frac{1}{\tau}\sum_{t=1}^\tau\E{r(t)}=\mu>\mu-\delta^-$, the function $(\mu-\delta^-)\theta+\bar\Lambda_{r}(-\theta)$ has a negative slope at $\theta=0$. Hence, we can choose some $\tilde{\theta}>0$, such that $(\mu-\delta^-)\tilde{\theta}+\bar\Lambda_{r}(-\tilde{\theta})<0$. This implies that there exists a $J$ and a $\beta>0$ such that for every $i>J$,
\begin{align}
f_i(\tilde{\theta})&>\gamma+i\beta
\end{align}
as illustrated in Fig.~\ref{fig:diffslopeproof}. Returning to Eq.~(\ref{eq:probsum2}),
\begin{align}
&p_\text{discharge}^{\mathcal{B}}(M) = \sum_{i=0}^\infty \Prob{E_i} \notag\\
&\leq \sum_{i=0}^\infty\Prob{\sum_{k=-(i+1)\lceil\frac{M}{2A}\rceil+1}^0(e^\mathcal{B}(k)-r(k))>\frac{M}{2}}\notag\\
&\leq\sum_{i=0}^J \exp\left(-\frac{M}{2}\left[f_i(\theta_i^*)+\epsilon_i(M,\theta_i^*)\right]\right)\notag\\
&\qquad+\sum_{i=J+1}^\infty \exp\left(-\frac{M}{2}\left[f_i(\tilde{\theta})+\epsilon_i(M,\tilde{\theta})\right]\right)\notag\\
&\leq \sum_{i=0}^J \exp\left(-\frac{M}{2}\left[\gamma+\min_{0\leq i\leq J}\epsilon(M,\theta_i^*)\right]\right)\notag\\
&\qquad+\sum_{i=J+1}^\infty \exp\left(-\frac{M}{2}\left[\gamma+i\beta+\inf_{i>J}\epsilon_i(M,\tilde\theta)\right]\right)\notag\\
&=\exp\left(-\frac{M}{2}\gamma\right)\Bigg[(J+1)\exp\left(\min_{0\leq i\leq J}\epsilon(M,\theta_i^*)\right)\notag\\
&\qquad+\frac{\exp\left(-\frac{M}{2}\left((J+1)\beta+\inf_{i>J}\epsilon_i(M,\tilde\theta)\right)\right)}{1-\exp\left(-\beta \frac{M}{2}\right)}\Bigg].\label{eq:bigeqn}
\end{align}
From Eq.~(\ref{eq:bigeqn}), $\limsup_{M\rightarrow\infty} \frac{2}{M}\log p_\text{discharge}^{\mathcal{B}}(M)\leq -{\gamma}$.
Since this inequality holds for any $A>0$, we let $A\rightarrow\infty$ as follows:
\begin{align}
&\limsup_{M\rightarrow\infty}\frac{2}{M}\log p_\text{discharge}^{\mathcal{B}}(M)\notag\\
&\quad\leq -\inf_{i\geq0}\sup_{\theta\geq0}\left[\theta\left(1-\frac{i}{A}(\mu-\delta)\right)-\frac{i}{A}\bar\Lambda_{r}(-\theta)\right]\notag\\
&\quad=-\inf_{T\geq0}\sup_{\theta\geq0}\left[\theta\left(1-T(\mu-\delta)\right)-T\bar\Lambda_{r}(-\theta)\right]\notag\\
&\quad=-\inf_{T\geq0}T\sup_{\theta\geq0}\left[-\theta\left(\mu-\delta-\frac{1}{T}\right)-\bar\Lambda_{r}(-\theta)\right]\label{eq:ub2}.
\end{align}

Next, we find the {lower} bound.
For some $T\geq0$, a sufficient condition for the battery to be empty at some time slot in the interval $[-\lceil TM/2 \rceil,0]$ is that,
\begin{align}
\sum_{t=-\lceil \frac{TM}{2} \rceil +1}^0 (e^\mathcal{B}(t)-r(t))&>M.
\end{align}
We can lower bound $p_\text{discharge}^{\mathcal{B}}(M)$ using the union bound,
\begin{align}
\nonumber
&\Prob{B(t)=0\ \text{within some}\ t\in \left[-\left\lceil \frac{TM}{2}\right\rceil,0\right]} \\
\nonumber
&=\Prob{\bigcup_{t=-\lceil TM/2 \rceil}^0 B(t) = 0} \leq \sum_{t=-\lceil TM/2 \rceil}^0 \Prob{B(t)=0} \\
\label{eq32}
&= \left\lceil \frac{TM}{2} \right\rceil p_\text{discharge}^{\mathcal{B}}(M).
\end{align}
We also have
\begin{align}
&\Prob{\sum_{t=-\lceil \frac{TM}{2} \rceil +1}^0 (e^\mathcal{B}(t)-r(t))>\frac{M}{2}}\notag\\
&\quad= \Prob{\sum_{t=-\lceil \frac{TM}{2} \rceil +1}^0 (\mu-\delta^--r(t))>\frac{M}{2}}\label{eq:lowerbound12}.
\end{align}
We define, $Z_{M,T}\triangleq\frac{2}{TM}\sum_{k=-\lceil \frac{TM}{2}\rceil+1}^0\left(\mu-\delta^--r(k)\right)$. Consequently,
\begin{align*}
\Prob{\sum_{t=-\lceil\frac{TM}{2} \rceil +1}^0 (\mu-\delta^--r(t))>\frac{M}{2}}&=\Prob{Z_{M,T}>\frac{1}{T}}.
\end{align*}
Now, $\lim_{M\rightarrow\infty}\E{Z_{M,T}}=-\delta^-<0<\frac{1}{T}$ for all $T>0$.
Applying the G\"artner-Ellis Theorem, we get,
\begin{align}
&\lim_{M\rightarrow\infty}\frac{2}{M}\log\Prob{Z_{M,T}>\frac{1}{T}}\notag\\
&\quad=-\sup_{s\geq0}\left[\frac{1}{T}s-s\left(\mu-\delta\right)+T\bar\Lambda_r\left(-\frac{s}{T}\right)\right]\notag\\
&\quad=-T\sup_{s\geq0}\left[-\frac{s}{T}\left(\mu-\delta-\frac{1}{T}\right)-\bar\Lambda_r\left(-\frac{s}{T}\right)\right]\notag\\
&\quad=-T\sup_{\theta\geq0}\left[-\theta\left(\mu-\delta-\frac{1}{T}\right)-\bar\Lambda_{r}(-\theta)\right].\label{eq22}
\end{align}
Combining Eqs.~(\ref{eq32}) and (\ref{eq22}), we have,
\begin{align}
&\liminf_{M\rightarrow\infty}\frac{2}{M}\log p_\text{discharge}^{\mathcal{B}}(M)\notag\\
&\quad\geq-\inf_{T\geq0}T\sup_{\theta\geq0}\left[-\theta\left(\mu-\delta-\frac{1}{T}\right)-\bar\Lambda_{r}(-\theta)\right]\label{eq:lb2}.
\end{align}

From Eqs.~(\ref{eq:ub2}) and~(\ref{eq:lb2}) we have,
\begin{align}
&\lim_{M\rightarrow\infty}\frac{2}{M}\log p_\text{discharge}^{\mathcal{B}}(M)\notag\\
&\quad=-\inf_{T\geq0}T\sup_{\theta\geq0}\left[-\theta\left(\mu-\delta-\frac{1}{T}\right)-\bar\Lambda_{r}(-\theta)\right]\notag\\
&\quad=s_{d^-}^*\label{eq:tight2}.
\end{align}
This gives us $p_\text{discharge}^{\mathcal{B}}(M)=\Theta\left(\exp\left(s_{d^-}^*\frac{M}{2}\right)\right)$.
\end{IEEEproof}

\begin{lemma}\label{lemma:s}
The asymptotic variance of $r(t)$, $\bar\sigma_r^2\triangleq\lim_{T\to\infty}\frac{1}{T}\var{\sum_{t=1}^Tr(t)}$ satisfies
\begin{align}
\left.\frac{\partial s_{d^-}^*}{\partial\delta^-}\right|_{\delta^-=0}=-\frac{2}{\bar\sigma_r^2}
\end{align}
\end{lemma}
\begin{IEEEproof}
First, we define $\bar\Lambda_{d^{-}}^{(n)}(0)=\left.\frac{\partial^n \bar\Lambda_{d^{-}}(s)}{\partial s^n}\right|_{s=0}$. The Taylor series expansion of $\bar\Lambda_{d^{-}}(s_{d^-}^*)$ about $s=0$ gives,
\begin{align*}
0=\bar\Lambda_{d^{-}}(s^*)&=\sum_{n=0}^\infty \bar\Lambda_{d^{-}}^{(n)}(0)\frac{(s_{d^-}^*)^n}{n!}\\
&=\underbrace{\bar\Lambda_{d^{-}}(0)}_{=0}+\bar\Lambda_{d^{-}}^{(1)}(0)s_{d^-}^*+\bar\Lambda_{d^{-}}^{(2)}(0)\frac{(s_{d^-}^*)^2}{2!}+\cdots\\
&=\sum_{n=1}^\infty \bar\Lambda_{r}^{(n)}(0)\frac{(s_{d^-}^*)^n}{n!}-(\mu-\delta^-)s_{d^-}^*\\
&=\mu s_{d^-}^*+\sum_{n=2}^\infty \bar\Lambda_{r}^{(n)}(0)\frac{(s_{d^-}^*)^n}{n!}-(\mu-\delta)s_{d^-}^*.
\end{align*}
Rearranging the terms, we have
\begin{align}
\sum_{n=2}^\infty \bar\Lambda_r^{(n)}(0)\frac{(s_{d^-}^*)^{n-1}}{n!} &= -\delta^-.
\end{align}
Differentiating with respect to $\delta^-$, we have,
\begin{align*}
\frac{\partial s_{d^-}^*}{\partial\delta^-}\sum_{n=2}^\infty \bar\Lambda_r^{(n)}(0)\frac{(n-1)(s_{d^-}^*)^{n-2}}{n!}=-1.
\end{align*}
As $\delta^-\rightarrow 0$, $s_{d^-}^*\rightarrow 0$ the above expression reduces to,
\begin{align}
\left.\frac{\partial s_{d^-}^*}{\partial\delta^-}\right|_{\delta^-=0} \bar\Lambda_r^{(2)}(0)\frac{1}{2}=-1.\label{eq:2ndmom}
\end{align}
Substituting $\bar\Lambda_r^{(2)}(0)=\bar\sigma_r^2$ in Eq.~(\ref{eq:2ndmom}), we have the required result.
\end{IEEEproof}

Lemma~\ref{lemma:s} implies $\frac{\partial s_{d^-}^*}{\partial\delta^-} = -\frac{2}{\bar\sigma_r^2}+\text{o}(\delta^-)$
and hence,
\begin{align}
s_{d^-}^*&=-\frac{2}{\bar\sigma_r^2}\delta^-+\text{o}(\delta^-),
\end{align}
where $\text{o}(\delta^-)/\delta^- \rightarrow 0$ as $\delta^- \rightarrow 0$.

Substituting this in Eq.~(\ref{eq:tight2}), we have,
\begin{align}
p_\text{discharge}^\mathcal{B}(M) &= \Theta\left(\exp\left[\left(-\frac{2}{\bar\sigma_r^2}\delta^- + \text{o}\left(\delta^-\right)\right)\frac{M}{2}\right]\right).
\end{align}
%
By choosing $\delta^-=\alpha\frac{\log M}{M}$ and $\alpha=\beta\bar\sigma_r^2$ we have $p_\text{discharge}^\mathcal{B}(M)=\Theta(M^{-\beta})$.

Next we show that with $\delta^+=\alpha\frac{\log M}{M}$ the scheme achieves an average utility $\bar U^\mathcal{B}$ such that $U(\mu)-\bar{U}^\mathcal{B}=\Theta\left(\frac{(\log M)^2}{M^2}\right)$. The instantaneous utility $U(e^\mathcal{S}(t))$ is zero with an $\Theta(M^{-\beta})$ probability. For the remaining time, the utility alternates between $U^+$ and $U^-$ as illustrated in Fig.~\ref{fig:utility_curve}. {Noting that $U(\cdot)$ is an analytic function on the non-negative real line}, the Taylor series expansion of the utility function about $\mu$ will be,
\begin{align}
U^+&=U(\mu)+U^{(1)}(\mu)\delta^++U^{(2)}(\mu)(\delta^+)^2+\text{o}((\delta^+)^2), \notag
\end{align}
and,
\begin{align}
U^-&=U(\mu)-U^{(1)}(\mu)\delta^-+U^{(2)}(\mu)(\delta^-)^2+\text{o}((\delta^-)^2). \notag
\end{align}
We define $\rho^+$ as the fraction of time that $B(t)>M/2$ and $\rho^-=1-\rho^+$ as the fraction of time that $B(t)\leq M/2$. The average utility $\bar{U}^{\mathcal{B}}$ can be written as,
\begin{align}
\bar{U}^{\mathcal{B}}&=\rho^+U^++(\rho^--p_\text{discharge}^\mathcal{B}(M))U^- \notag\\
&=U(\mu)+U^{(1)}(\mu)(\rho^+\delta^+-\rho^-\delta^-)+\Theta\left(\frac{(\log M)^2}{M^2}\right),
\label{eq:genutilexp}
\end{align}
where Eq.~(\ref{eq:genutilexp}) follows from the fact that $\delta^-,\delta^+=\alpha\frac{\log M}{M}$ and $p_\text{discharge}^\mathcal{B}(M)=\Theta(M^{-\beta})$ where $\beta\geq2$.

From conservation of energy, the replenishment energy is consumed completely except for the amount lost due to battery overflows. Thus,
\begin{align}
&\rho^+(\mu+\delta^+)+(\rho^--p_\text{discharge}^\mathcal{B}(M))(\mu-\delta^-)\notag\\
&\qquad\qquad\qquad\qquad\qquad\qquad=\mu(1-p_\text{overflow}^\mathcal{B}(M)),\label{eq:flow}
\end{align}
where $p_\text{overflow}^\mathcal{B}(M)$ is the probability of the battery being full under the energy management scheme $\mathcal{B}$. By a trivial extension of Lemmas~\ref{lemma:exprate2} and~\ref{lemma:s}, it can be shown that $p_\text{overflow}^\mathcal{B}(M)=\Theta\left(M^{-\beta}\right)$. We can simplify Eq.~(\ref{eq:flow}) as,
\begin{align}
\rho^+\delta^+-\rho^-\delta^-&=\Theta\left(M^{-\beta}\right).\label{eq:firstorder}
\end{align}
By substituting Eq.~(\ref{eq:firstorder}) in the first-order term of Eq.~(\ref{eq:genutilexp}), we observe that the scheme $\mathcal{B}$ achieves $U(\mu)-\bar{U}^\mathcal{B}=\Theta\left(\frac{(\log M)^2}{M^2}\right)$. Choosing $\delta^\mathcal{B}=\delta^+=\alpha\frac{\log M}{M}$ in Eq.~(\ref{allocation}) completes the proof of Theorem~\ref{th:quaddecay}.

\section{Proof of Theorem~\ref{th:expdecay}}\label{app:prop1}
Consider any ergodic energy management scheme $\mathcal{S}$ that uses $e^\mathcal{S}(t)$ units of energy in the time slot $t$. Note that scheme $\mathcal{S}$ can be deterministic or randomized. The asymptotic semi-invariant log moment generating function of the net battery drift $d^\mathcal{S}(t)\triangleq e^\mathcal{S}(t)-r(t)$ is given by $\bar\Lambda_{d^\mathcal{S}}(s)$.
First, we state a lemma that gives the discharge probability scaling for scheme $\mathcal{S}$.
\ignore{Since this lemma is a minor modification of Lemma~\ref{lemma:exprate2}, we omit the proof in this paper. We direct the reader to~\cite{techrep2} for the detailed proof of this lemma.}
\begin{lemma}\label{lemma:exprate}
The probability of battery discharge under Scheme $\mathcal{S}$ with battery size $M$ follows $p_\text{discharge}^{\mathcal{S}}(M)=\Theta(\exp(-s_{d^\mathcal{S}}^*M))$, where $s_{d^\mathcal{S}}^*$ is the positive root of $\bar{\mu}_{d^\mathcal{S}}(s)$.
\end{lemma}
\begin{IEEEproof}
This lemma gives the rate of decay of the probability of complete discharge with respect to the battery size $M$. To prove this result, we first find an \emph{upper bound} for the required probability. We fix a constant $A>0$ and decompose the time line into intervals, such that each interval is of length $\lceil \frac{M}{A}\rceil$ and the $i$th interval ends at time slot $t_i=i\lceil\frac{M}{A}\rceil$. Next, define $E_i$ as the event that the battery is empty at the end of time slot 0 and the last time the battery was full (i.e.,~$M$) is some time during the interval $-i=\left[-(i+1)\lceil\frac{M}{A}\rceil+1,-i\lceil\frac{M}{A}\rceil\right]$. The event of an empty battery at time slot 0 can be decomposed as a union of events $E_i$,
\begin{align}
p_\text{discharge}^{\mathcal{S}}(M) &= \sum_{i=0}^\infty \Prob{E_i}\label{eq:probsum}
\end{align}

A necessary condition for event $E_i$ to occur is,
\begin{align}
\sum_{t=-(i+1)\lceil\frac{M}{A}\rceil+1}^0 \left(e^\mathcal{S}(t)-r(t)\right)>M.
\end{align}
Using Chernoff's bound, for any $\theta_i\geq0$,
\begin{align}
&\Prob{\sum_{t=-(i+1)\lceil\frac{M}{A}\rceil+1}^0d^\mathcal{S}(t)>M} \notag\\
&\leq\E{\exp\left(\theta_i\sum_{t=-(i+1)\lceil\frac{M}{A}\rceil+1}^0d^\mathcal{S}(t)\right)}\exp\left(-\theta_iM\right) \notag\\
&=\exp\left(-{M}\left(\theta_i-\frac{i+1}{A}\bar\Lambda_{d^\mathcal{S}}(\theta_i)+\epsilon_i(M,\theta_i)\right)\right),
\end{align} 
where $\epsilon_i(M,\theta_i)\rightarrow0$ as $M\rightarrow\infty$. 

In order to find the tightest bound for each $i$, we choose $\theta_i^*\geq 0$ to maximize $f_i(\theta)\triangleq\theta-\frac{i+1}{A}\bar\Lambda_{d^\mathcal{S}}(\theta)$,
and define $\gamma=\inf_{i\geq0}\sup_{\theta\geq0}f_i(\theta)$.
We can rewrite $f_i(\theta)$ as,
\begin{align}
f_i(\theta)&=\theta-\frac{\bar\Lambda_{d^\mathcal{S}}(\theta)}{A}-i\frac{\bar\Lambda_{d^\mathcal{S}}({\theta})}{A}.\notag
\end{align}
\ignore{\begin{figure}
\centering
\psfrag{0}[cc]{$0$}
\psfrag{i}[cc]{$i$}
\psfrag{K}[cc]{$J$}
\psfrag{a}[lc]{$\gamma+i\beta$}
\psfrag{c}[cc]{$\gamma$}
\psfrag{b}[lc]{$f_i(\tilde{\theta})$}
\psfrag{0}[cc]{$0$}
\psfrag{slope}[cc]{slope=$\text{E}\left[X_i\right]=-\delta_1^{(a)}$}
\psfrag{min1}[rc]{$\inf_{r\leq r_{\text{buffer}}^*}\mu(r)$}
\psfrag{min2}[lc]{$-\mu_{r_{\text{buffer}}^*}^*(0)$}
\includegraphics[width=0.3\textwidth]{figures/diffslopes.eps}
\caption{A geometric proof for the existence of $K$ and $\delta>0$ such that for every $i>K$, $f_i(\tilde{\theta})>\gamma+i\delta$.} \label{fig:diffslopeproof}
\end{figure}}

Since $\lim_{\tau\to\infty}\E{d^\mathcal{S}(\tau)}<0$, the function $\bar\Lambda_{d^\mathcal{S}}(\theta)$ has a negative slope at $\theta=0$. Hence, we can choose some $\tilde{\theta}>0$, such that $\bar\Lambda_{d^\mathcal{S}}(\tilde{\theta})<0$. This implies that there exists a $J$ and a $\beta>0$ such that for every $i>J$ (see Fig.~\ref{fig:diffslopeproof} for a graphical proof),
\begin{align}
f_i(\tilde{\theta})&>\gamma+i\beta.
\end{align}

Returning to Eq.~(\ref{eq:probsum}),
\begin{align}
&p_\text{discharge}^{\mathcal{S}}(M) = \sum_{i=0}^\infty \Prob{E_i} \notag\\
&\leq \sum_{i=0}^\infty\Prob{\sum_{t=-(i+1)\lceil\frac{M}{A}\rceil+1}^0d^\mathcal{S}(t)>M}\notag\\
&\leq\sum_{i=0}^J \exp\left(-{M}\left( f_i(\theta_i^*)+\epsilon_i(M,\theta_i^*)\right)\right)\notag\\
&\qquad+\sum_{i=J+1}^\infty \exp\left(-{M}\left(f_i(\tilde{\theta})+\epsilon_i(M,\tilde{\theta})\right)\right)\notag\\
&\leq \sum_{i=0}^J \exp\left(-{M}\left(\gamma+\min_{0\leq i\leq J}\epsilon(M,\theta_i^*)\right)\right)\notag\\
&\qquad+\sum_{i=J+1}^\infty \exp\left(-{M}\left(\gamma+i\beta+\inf_{i>J}\epsilon_i(M,\tilde\theta)\right)\right)\notag\\
&=\exp\left(-{M}\gamma\right)\Bigg[(J+1)\exp\left(\min_{0\leq i\leq J}\epsilon(M,\theta_i^*)\right)\notag\\
&\qquad+\frac{\exp\left(-{M}\left((J+1)\beta+\inf_{i>J}\epsilon_i(M,\tilde\theta)\right)\right)}{1-\exp\left(-\beta M\right)}\Bigg].
\end{align}
As $M\rightarrow\infty$,
\begin{align}
&\limsup_{M\rightarrow\infty} \frac{1}{M}\log p_\text{discharge}^{\mathcal{S}}(M)\leq -{\gamma}\notag\\
\end{align}
Since this inequality holds for any $A>0$, we let $A\rightarrow\infty$ and get,
\begin{align}
\limsup_{M\rightarrow\infty}\frac{1}{M}\log p_\text{discharge}^{\mathcal{S}}(M)&\leq -\inf_{i\geq0}\sup_{\theta\geq0}\left[\theta-\frac{i}{A}\bar\Lambda_{d^\mathcal{S}}(\theta)\right]\notag\\
&=-\inf_{T\geq0}\sup_{\theta\geq0}\left[\theta-T\bar\Lambda_{d^\mathcal{S}}(\theta)\right]\notag\\
&=-\inf_{T\geq0}T\sup_{\theta\geq0}\left[\frac{\theta}{T}-\bar\Lambda_{d^\mathcal{S}}(\theta)\right].\label{eq:ub}
\end{align}

Next, we find the \emph{lower bound}.
For some $T\geq0$, a sufficient condition for the battery to be empty at some time slot in the interval $[-\lceil TM \rceil,0]$ is that,
\begin{align}
\sum_{t=-\lceil TM \rceil +1}^0 (e^\mathcal{S}(t)-r(t))&>M.
\end{align}
We can lower bound $p_\text{discharge}^{\mathcal{S}}(M)$ using the union bound,
\begin{align}
&p_\text{discharge}^{\mathcal{S}}(M) \lceil TM \rceil \notag\\
&\geq \Prob{\text{battery is empty in some slot during $[-\lceil TM\rceil,0]$}}.\label{eq3}
\end{align}
We define, $Z_{M,T}\triangleq\frac{1}{TM}\sum_{t=-\lceil TM\rceil+1}^0d^\mathcal{S}(t)$. Consequently,
\begin{align*}
\Prob{\sum_{t=-\lceil TM \rceil +1}^0 d^\mathcal{S}(t)>{M}}&=\Prob{Z_{M,T}>\frac{1}{T}}.
\end{align*}
Now, $\lim_{M\rightarrow\infty}\E{Z_{M,T}}<0<\frac{1}{T}$ for all $T>0$. 
\ignore{The asymptotic semi-invariant moment generating function of $Z_{M,T}$ can be written as,
\begin{align}
\lim_{M\rightarrow\infty}\frac{1}{M}\log\E{\exp\left\{sMZ_{M,T}\right\}}&=\lim_{M\rightarrow\infty}\frac{1}{M}\log\E{\exp\left\{sM\frac{1}{TM}\sum_{t=-\lceil TM\rceil+1}^0\left(\mu-\delta_1^{(r)}-R_t\right)\right\}}\notag\\
&=s\left(\mu^{(r)}-\delta^{(r)}\right)+\frac{1}{M}\log\E{\exp\left\{-\frac{s}{T}\sum_{t=-\lceil TM\rceil+1}^0R_t\right\}}\notag\\
&=s\left(\mu^{(r)}-\delta^{(r)}\right)+T\mu^{(r)}\left(-\frac{s}{T}\right).
\end{align}}
Applying the G\"artner-Ellis Theorem, we get,
\begin{align}
\lim_{M\rightarrow\infty}\frac{1}{M}\log\Prob{Z_{M,T}>\frac{1}{T}}&=-\sup_{s\geq0}\left[\frac{s}{T}-\bar\Lambda_{Z}
\left(s\right)\right]\notag\\
&=-T\sup_{s\geq0}\left[\frac{s}{T^2}-\bar\Lambda_{d^\mathcal{S}}\left(\frac{s}{T}\right)\right]\notag\\
&=-T\sup_{\theta\geq0}\left[\frac{\theta}{T}-\bar\Lambda_{d^\mathcal{S}}(\theta)\right],\label{eq2}
\end{align}
where $\bar\Lambda_Z(s)$ is the asymptotic semi-invariant log-moment generating function of $Z_{M,T}$.
By noting that Eq.~(\ref{eq2}) holds for all $T\geq0$ and combining it with Eq.~(\ref{eq3}), we have,
\begin{align}
\liminf_{M\rightarrow\infty}\frac{1}{M}\log p_\text{discharge}^{\mathcal{S}}(M)&\geq-\inf_{T\geq0}T\sup_{\theta\geq0}\left[\frac{\theta}{T}-\bar\Lambda_{d^\mathcal{S}}(\theta)\right]\label{eq:lb}.
\end{align}

From Eqs.~(\ref{eq:ub}) and~(\ref{eq:lb}) we have,
\begin{align}
\lim_{M\rightarrow\infty}\frac{1}{M}\log p_\text{discharge}^{\mathcal{S}}(M)&=
-\inf_{T\geq0}T\sup_{\theta\geq0}\left[\frac{\theta}{T}-\bar\Lambda_{d^\mathcal{S}}(\theta)\right]\notag\\
&=-s_{d^\mathcal{S}}^*\label{eq:tight}.
\end{align}
This gives us $p_\text{discharge}^{\mathcal{S}}(M)=\Theta(\exp(-s_{d^\mathcal{S}}^*M))$.
\end{IEEEproof}

Note that $s_{d^\mathcal{S}}^*>0$ exists\footnote{Since $\bar\Lambda_{d^\mathcal{S}}(0)$ and $\left.\frac{\partial \bar{\Lambda}_{d^\mathcal{S}}(s)}{\partial s}\right|_{s=0}=\lim_{\tau\to\infty}\E{d^\mathcal{S}(\tau)}<0$, $s_{d^\mathcal{S}}^*>0$ will exist.} if and only if $\E{d^\mathcal{S}(t)}<0$. Therefore, for $p_\text{discharge}^\mathcal{S}(M)$ to decay exponentially with $M$, we require,
\begin{align}
\E{e^\mathcal{S}(t)}<\E{r(t)}=\mu.\label{eq:th2ineq1}
\end{align}
On the other hand, if $\E{d^\mathcal{S}(t)}\geq0$, there exists no rate $s>0$ at which the battery discharge probability decays exponentially with $M$, i.e., $p_\text{discharge}^\mathcal{S}(M) = \Omega (\exp(-sM))$ for all $s>0$. By substituting $\alpha_c=s_{d^\mathcal{S}}^*$ in Lemma~\ref{lemma:exprate}, we get the required scaling law $p_\text{discharge}^\mathcal{S}(M)=\Theta(\exp(-\alpha_cM))$.

The difference between the utilities is given by,
{\begin{align}
U(\mu)-\bar{U}^\mathcal{S}&=U(\mu)-\liminf_{\tau\to\infty}\frac{1}{\tau}\sum_{t=1}^\tau{U(e^\mathcal{S}(t))}\notag\\
&\stackrel{(a)}{\geq} U(\mu)-U\left(\liminf_{\tau\to\infty}\frac{1}{\tau}\sum_{t=1}^\tau{e^\mathcal{S}(t)}\right)\notag\\
&\stackrel{(b)}{=}U(\mu)-U\left(\E{e^\mathcal{S}(t)}\right)\stackrel{(c)}{=}\Omega(1).\label{eq:Jensen2}
\end{align}}
Where $(a)$ follows from Eq.~(\ref{eq:infjensen}), $(b)$ is due to the ergodicity of $e^\mathcal{S}(t)$ and $(c)$ follows from Eq.~(\ref{eq:th2ineq1}) and the fact $U(\cdot)$ is an increasing function.
This completes the proof for Theorem~\ref{th:expdecay}.

\section{Proof of Theorem~\ref{th:bufferbattery}}\label{app:BrownianProof}
{As discussed in Section~\ref{subsec:asyoptimal}, our proof is constructive. We use the energy management scheme $\mathcal{Q}$ given in Eq.~(\ref{joint_allocation}).  
We find the individual probabilities in the following lemmas.
\begin{lemma}\label{lemma:QBufferOverflow}
For the energy management scheme $\mathcal{Q}$, given any $\beta_Q \geq 2$, $p_\text{overflow}^{\cal Q}(K)=\text{O}(K^{-\beta_Q})$.
\end{lemma}
\begin{IEEEproof}
{
\ignore{The time instants $\{T_n,\;n=1,2,\ldots\}$ at which the buffer process returns to state $K/2$ are iteratively defined as $T_i\triangleq \argmin_{t>T_{i-1}}\{Q(t)=K/2\}$ and $T_0=0$. We introduce an indicator function, $\Delta_i^u$,
\begin{align}
\Delta_i^{u} =
\begin{cases}
1, &\text{if}\ Q(t)>K/2, \ \forall \ t \ \in \ (T_{i-1}, T_i)\\
0, &\text{if}\ Q(t)<K/2, \ \forall \ t \ \in \ (T_{i-1}, T_i)
\end{cases}.
\end{align}}
\begin{figure}
\centering
\psfrag{q}[cb]{$Q(t)$}
\psfrag{k}[rc]{$\frac{K}{2}$}
\psfrag{u}[cb]{$Q_u(t)$}
\psfrag{t}[lc]{$t$}
\psfrag{s}[lc]{$t$}
\psfrag{b}[lc]{$f_i(\tilde{\theta})$}
\psfrag{0}[cc]{$0$}
\psfrag{slope}[cc]{slope=$\text{E}\left[X_i\right]=-\delta_1^{(a)}$}
\psfrag{min1}[rc]{$\inf_{r\leq r_{\text{buffer}}^*}\mu(r)$}
\psfrag{min2}[lc]{$-\mu_{r_{\text{buffer}}^*}^*(0)$}
\includegraphics[height=1.1in]{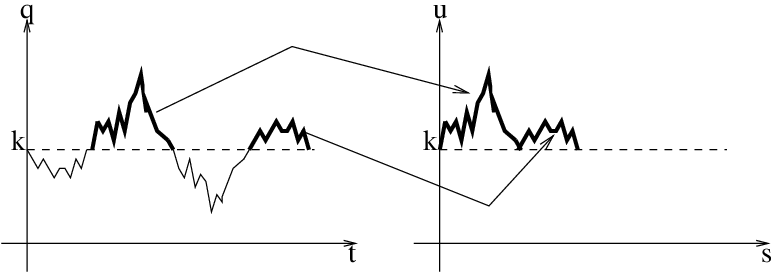}
\caption{A graphical representation of the relationship between $Q(t)$ and $Q_u(t)$.} \label{fig:qvsqu}
\vspace*{-0.2in}
\end{figure}
First, consider a process $Q_u(t)$ that is formed by splicing together the intervals during which the process $Q(t)$ is above the buffer state $K/2$. Fig.~\ref{fig:qvsqu} illustrates a sample path of the process $Q(t)$ and the corresponding $Q_u(t)$.
\ignore{\begin{align*}
&Q(t)\\
&=\begin{dcases}
Q_u\left(\Delta_{1}^ut\right), & t\leq T_1\\
Q_u\left(\sum_{i=1}^{j} \Delta_i^u(T_{i}-T_{i-1})+\Delta_{j+1}^u(t-T_{j})\right) ,& t>T_1
\end{dcases}
\end{align*}
where $j=\argmax_{k>0}\left(\sum_{i=1}^k \Delta_i^u(T_i-T_{i-1})<t\right)$.
\ignore{
\begin{align}
j(t) = 
\begin{cases}
0, & t\leq T_1\\
\underset{k>0}\argmax\left(\sum_{i=1}^k \Delta_i^u(T_i-T_{i-1})<t\right)& t> T_1
\end{cases}.
\end{align}}}
We denote the diffusion limit of $Q(t)$ and $Q_u(t)$ by $\mathbf{Q}(t)$ and $\mathbf{Q}_u(t)$, respectively. 
While $\mathbf{Q}(t)$ is not a Brownian motion due to its state-dependent drift, $\mathbf{Q}_u(t)$ will be a Brownian motion in the large-battery regime, with reflections at $K/2$ and unbounded from above as detailed in Step (T1).
We assume the starting state of $\mathbf{Q}_u(t)$ to be $\mathbf{Q}_u(0)=K/2$. 
Note that due to the strong Markovian property of a Brownian motion~\cite{HarrisonBrownian}, the instants $\{T_n^u,\;n=1,2,\ldots\}$ at which the process $\mathbf{Q}_u(t)$ returns to state $K/2$ (i.e.,~$\mathbf{Q}_u(T_i^u)=K/2$) is probabilistically equal to the starting state. Hence we can study these renewal epochs\footnote{If we assume the starting state to be $\mathbf{Q}_u(0)\neq K/2$, we can simply consider the process to be a delayed renewal process. The steady state properties in the resulting analysis will not change.} to obtain steady state properties for the data queue process. 
\ignore{Now, consider the random variables { $T_1^u\triangleq\arg\min_{t>0}\{\mathbf{Q}_u(t)=K/2\;|\;\mathbf{Q}_u(0)=K/2\}$ and $T_1^l\triangleq\arg\min_{t>0}\{\mathbf{Q}_l(t)=K/2\;|\;\mathbf{Q}_l(0)=K/2\}$}. The hitting time distributions can be calculated as~\cite{HarrisonBrownian},
\begin{align}
&\Prob{T_1^u>t+\tau\left|\mathbf{Q}_u(\tau)=\frac{K}{2}+\epsilon\right.}\notag\\
&\quad=\Phi\left(\frac{\epsilon-\delta^{(a)}\tau}{\bar\sigma_a\sqrt{\tau}}\right)-\exp\left(\frac{2\delta^{(a)}\epsilon}{\bar\sigma_a^2}\right)\Phi\left(\frac{-\epsilon-\delta^{(a)}\tau}{\bar\sigma_a\sqrt{\tau}}\right)\notag\\
&\quad=\Prob{T_1^l>t+\tau\left|\mathbf{Q}_l(\tau)=\frac{K}{2}-\epsilon\right.},\label{eq:BMComp}
\end{align}
where $\Phi(y)\triangleq\frac{1}{\sqrt{2\pi}}\int_{-\infty}^y \exp\left(-\frac{x^2}{2}\right)dx$. Since Eq.~(\ref{eq:BMComp}) holds for all $\tau>0$ and $\epsilon>0$, the random variables $T_1^u$ and $T_1^l$ will have the same distribution. Furthermore, once the process $\mathbf{Q}(t)=K/2$, it can go above or below $K/2$ with equal probability. Consequently, we need to study the renewals associated with $\mathbf{Q}_u(t)$ and can find $\Prob{\mathbf{Q}(t)>K}=\frac{1}{2}\Prob{\mathbf{Q}_u(t)>K}$, which is identical to $p_\text{overflow}^{\cal Q}(K)$.}}

If we define a unit reward (i.e.,~$R(t)=1$) for every time $t$ that the process $\mathbf{Q}_u(t)>K$ then,
\begin{align}
\lim_{t\rightarrow\infty}P(\mathbf{Q}_u(t)>K)&=\lim_{t\rightarrow\infty}\E{R(t)}.
\end{align}
From renewal-reward theory~\cite{HarrisonBrownian} we can write,
\begin{align}
\lim_{t\to\infty}\E{R(t)}=\frac{\E{R_n}}{\E{X}}\label{eq:reward},
\end{align}
where $\E{R_n}$ is the expected award accumulated in one renewal period, and $\E{X}$ is the expected length of the renewal period. To get the correct expression for $\lim_{t\to\infty}\E{R(t)}$, we need to write the expressions for $\E{R_n}$ and $\E{X}$ carefully. We define $\E{X(\epsilon)}$ as the expected time for process $\mathbf{Q}_u(t)$ to return to $K/2$ given that it starts at $K/2+\epsilon$.
The expression for $\E{X(\epsilon)}$ is given by~\cite{RossStochasticBook},
\begin{align}
\E{X(\epsilon)}=\frac{\epsilon}{\delta^{({a})}}.\label{eq:EX}
\end{align}
Similarily, we define $\E{R_n(\epsilon)}$ as the probability of reaching $K$ before $K/2$ starting at $K/2+\epsilon$. Passing the limit $\epsilon\downarrow0$ will give the expected reward accumulated in one renewal period. Applying the expression for this probability from~\cite{RossStochasticBook},
\begin{align}
\E{R_n(\epsilon)}&=\frac{\exp\left(\frac{2\delta^{({a})}}{\bar\sigma_a^2}\epsilon\right)-1}{\exp\left(\frac{2\delta^{({a})}}{\bar\sigma_a^2}\frac{K}{2}\right)-1}=\frac{\frac{2\delta^{({a})}}{\bar\sigma_a^2}\epsilon+o(\epsilon)}{\exp\left(\frac{2\delta^{({a})}}{\bar\sigma_a^2}\frac{K}{2}\right)-1}.\label{eq:ERn}
\end{align}
Dividing Eqs.~(\ref{eq:ERn}) by~(\ref{eq:EX}) and passing the limit $\epsilon\downarrow0$, we have,
\begin{align}
\lim_{t\rightarrow\infty}\E{R(t)}=\lim_{\epsilon\downarrow0}\frac{\left(\frac{2\delta^{({a})}}{\bar\sigma_a^2}+\frac{o(\epsilon)}{\epsilon}\right)\delta^{({a})}}{{\exp\left(\frac{2\delta^{({a})}}{\bar\sigma_a^2}\frac{K}{2}\right)-1}}.
\end{align}
Evaluating the limit $\epsilon\downarrow 0$, and noting that the overflow probability for the process $\mathcal{Q}_u(t)$ will be an upper bound on the overflow probability of process $\mathcal{Q}(t)$, for large $K$ we have,
\begin{align}
p_\text{overflow}^\mathcal{Q}(K)&\leq \lim_{t\rightarrow\infty}P(\mathbf{Q}_u(t)>K)\notag\\
&= \frac{2{\delta^{({a})}}^2}{\bar\sigma_a^2}\exp\left(-\frac{\delta^{({a})}K}{\bar\sigma_a^2}\right).
\end{align}
By choosing $\delta^{(a)}=\beta_Q \bar\sigma_a^2\frac{\log{K}}{K}$, we have,
\begin{align}
p_\text{overflow}^\mathcal{Q}(K)&\leq{\beta_Q}^2\bar\sigma_a^2\left(\frac{\log K}{K}\right)^2\exp\left(-\beta_Q \log K \right) \notag \\
&=\text{O}\left(K^{-\beta_Q}\right).\label{eq:overflow}
\end{align}
\end{IEEEproof}
{Note that Lemma~\ref{lemma:QBufferOverflow} implies that given any $\beta_Q>2$, there exists an energy management scheme $\mathcal{Q}$ that achieves an overflow probability $p_\text{overflow}^\mathcal{Q}(K)=\text{O}(K^{-\beta_Q})$. Following the discussion in Step (T1), this implies that for some $\beta>0$, $p_\text{loss}^\mathcal{Q}(K)=\text{O}(K^{-\beta})$.}

\begin{lemma}\label{lemma:QBatteryUnderflow}
For the energy management scheme $\mathcal{Q}$, 
$\lim_{\lambda\uparrow C(\mu)}\ \lim_{M\to\infty}\frac{1}{M(\mu-C^{-1}(\lambda))}\log(p_\text{underflow}^\mathcal{Q}(M)) \leq -\frac{2}{\bar\sigma_r^2}$, where $C^{-1}(\cdot)$ is the inverse of the analytic rate-power function $C(\cdot)$.
\end{lemma}
\begin{IEEEproof}
{Recall that $\mathbf{B}(t)$ is not a Brownian motion, but merely the process obtained by applying the diffusion limit on the battery process $B(t)$. In order to evaluate $p_\text{underflow}^\mathcal{Q}(M)=\lim_{t\to\infty}P(\mathbf{B}(t)<0)$, for scheme $\mathcal{Q}$, we introduce a new energy management scheme $\mathcal{Q}_l$, which allocates an amount of energy
identical to:
$e^{\mathcal{Q}_l}(t)=\mu-\delta_2^{(r)}$ 
for all $t$. Further, even when $Q(t)=0$, Scheme ${\cal Q}_l$ uses up energy $\mu-\delta_2^{(r)}$ to transmit dummy bits. Thus, the drift of the associated battery state is a constant, completely independent of the queue state. Hence, the diffusion limit for the associated battery process
yields a reflected Brownian motion with a single barrier at $M$ (recall that the lower barrier was removed). We denote this Brownian limit by $\mathbf{B}_l(t)$. 
For the same energy replenishment process $\{r(t), t\geq0\}$, the net battery drift is defined as $d^\mathcal{S}(t)\triangleq r(t)-e^\mathcal{S}(t)$, for a given scheme $\mathcal{S}\in\{\mathcal{Q}, \mathcal{Q}_l \}$. We know for all sample paths that, the net drifts satisfy:
\begin{align}
d^{\mathcal{Q}}(t)\geq d^{\mathcal{Q}_l}(t)
\label{eq:ineq_drift}
\end{align}
for all $t$.
The battery underflow probability for scheme $\mathcal{Q}_l$
is given by $p_\text{underflow}^{\mathcal{Q}_l}(M)=\lim_{t\to\infty}P(\mathbf{B}_l(t)<0)$.
}
It follows from Eq.~(\ref{eq:ineq_drift}) that,
\begin{align}
p_\text{underflow}^\mathcal{Q}(M)\leq p_\text{underflow}^{\mathcal{Q}_l}(M).\label{eq:squeeze}
\end{align}
{The Brownian limit $\mathbf{B}_l(t)$
is an exponentially distributed random variable}, and the underflow probability is given by~\cite{AsmussenQueueBook},
\begin{align}
p_\text{underflow}^{\mathcal{Q}_l}(M) &= \exp\left(-\frac{2\delta_2^{(r)}}{\bar\sigma_r^2} M \right),\label{eq:lefteq}
\end{align}
Note that, from the Taylor series expansion of the {analytic} rate-power function $C(\cdot)$, we have,
\begin{align}
\delta_2^{(r)}&=\mu-\left(C^{-1}(\lambda)+(C^{-1}(\lambda))^{(1)}\delta^{(a)}+\text{o}(\delta^{(a)})\right).
\end{align}
Since $\delta^{(a)}=\beta_Q\bar\sigma_a^2\frac{\log K}{K}\to 0$ as $K\to\infty$, we have $\lim_{K\to\infty} \delta_2^{(r)} = \mu-C^{-1}(\lambda)$ Using this observation and combining Eqs.~(\ref{eq:squeeze}) and~(\ref{eq:lefteq}) we have,
\begin{align}
\lim_{M\to\infty}\ \lim_{\lambda\uparrow C(\mu)}\frac{1}{M(\mu-C^{-1}(\lambda))}\log(p_\text{underflow}^\mathcal{Q}(M)) \leq -\frac{2}{\bar\sigma_r^2},
\label{eq:squeeze2}
\end{align}
completing the proof.
Note that, we have $\alpha_Q = \frac{2(\mu-C^{-1}(\lambda))}{\bar\sigma_r^2}$ in the heavy traffic limit.
\end{IEEEproof}

\ignore{To illustrate Lemma~\ref{lemma:QBatteryUnderflow}, we consider the Gaussian channel capacity given in Eq.~(\ref{channel_capacity}). We have, $\delta_1^{(r)}=\mu-\frac{\exp\left(\left(\lambda+\delta^{(a)}\right)\log 2\right)-1}{\gamma}$ and $\delta_2^{(r)}=\mu-\frac{\exp\left(\left(\lambda-\delta^{(a)}\right)\log 2\right)-1}{\gamma}$. We can use the power series expansion of the exponential function to get,
\begin{align}
\delta_1^{(r)}&=\mu-\frac{1}{\gamma}\left((\lambda+\delta^{(a)})\log 2+\frac{(\lambda+\delta^{(a)})^2(\log2)^2}{2}+\cdots\right)\notag\\
&=\mu-\frac{1}{\gamma}\exp(\lambda\log 2)+\Theta(\delta^{(a)}),
\end{align}
and,
\begin{align}
\delta_2^{(r)}&=\mu-\frac{1}{\gamma}\left((\lambda-\delta^{(a)})\log 2+\frac{(\lambda-\delta^{(a)})^2(\log2)^2}{2}+\cdots\right)\notag\\
&=\mu-\frac{1}{\gamma}\exp(\lambda\log 2)+\Theta(\delta^{(a)}).
\end{align}
Substituting these expressions in Eq.~(\ref{eq:squeeze2}), we get $p_\text{discharge}^\mathcal{Q}(M)=\text{O}(\exp(-\alpha_QM))$.}

Finally, we focus on the average utility $\bar{U}_D^\mathcal{Q}$ of the energy management scheme $\mathcal{Q}$. The instantaneous utility will be zero when the queue is empty or when the battery is discharged. 
\ignore{Due to symmetry of the Brownian limits, the empty buffer probability will be the same as the data loss 
\ignore{(i.e., full buffer)}probability.} Since $p_\text{discharge}^\mathcal{Q}(M)=\text{O}(\exp(-\alpha_QM))$, the contribution of the discharge term can be ignored, since it is a 0-probability event under the large battery regime. For the scheme $\mathcal{Q}$, the calculation of the average utility becomes exactly the same problem as that for the energy management scheme $\mathcal{B}$, which was analyzed in Appendix~\ref{app:prop2}. Here, we replace $B(t)$ with $Q(t)$ and the battery size $M$ with the data buffer size $K$ to get,
\begin{align}
U_D(\lambda)-\bar{U}_D^\mathcal{Q}&=\Theta\left(\frac{(\log K)^2}{K^2}\right).
\end{align}

{We repeat the proof here for completeness.
Noting that $U_D(\cdot)$ is an analytic function on the non-negative real line, the Taylor series expansion of the utility function about $\lambda$ will be,
\begin{align}
U_D^+&=U_D(\mu)+U_D^{(1)}(\mu)\delta^{(a)}+U_D^{(2)}(\mu)(\delta^{(a)})^2+\text{o}((\delta^{(a)})^2), \notag
\end{align}
and,
\begin{align}
U_D^-&=U_D(\mu)-U_D^{(1)}(\mu)\delta^{(a)}+U^{(2)}(\mu)(\delta^{(a)})^2+\text{o}((\delta^{(a)})^2). \notag
\end{align}
We define $\rho^+$ as the fraction of time that $Q(t)>K/2$ and $\rho^-=1-\rho^+$ as the fraction of time that $Q(t)\leq K/2$. The average utility $\bar{U}_D^\mathcal{Q}$ can be written as,
\begin{align}
\bar{U}_D^\mathcal{Q}&=\rho^+U_D^++(\rho^--p_\text{empty}^\mathcal{Q}(K))U_D^- \notag\\
&=U_D(\lambda)+U_D^{(1)}(\lambda)(\rho^+\delta^{(a)}-\rho^-\delta^{(a)})+\Theta\left(\frac{(\log K)^2}{K^2}\right),
\label{eq:genutilthm3}
\end{align}
where $p_\text{empty}^\mathcal{Q}(K)$ is the probability of the data buffer being empty under the scheme $\mathcal{Q}$. Eq.~(\ref{eq:genutilthm3}) follows from the fact that $\delta^{(a)}=\beta_Q \bar \sigma_a^2\frac{\log{K}}{K}$, while Lemmas~\ref{lemma:exprate2} and~\ref{lemma:s} yield $p_\text{empty}^\mathcal{Q}(K)=\Theta(K^{-\beta_Q})$ where $\beta_Q\geq2$.

{The average utility $\bar{U}_D^\mathcal{Q}$ can be written as,
\begin{align}
\bar{U}_D^\mathcal{Q}&=\frac{1}{2}\left(U_D(\lambda+\delta^{(a)})+U_D(\lambda-\delta^{(a)})\right)(1-p_\text{loss}^\mathcal{Q}(K))\notag\\
&=U_D(\lambda)+U_D^{(2)}(\lambda)(\delta^{(a)})^2+\text{o}\left((\delta^{(a)})^2\right)\label{eq:utilconv1}\\
&=U_D(\lambda)+\Theta\left(\beta_Q^2\frac{(\log K)^2}{K^2}\right), \label{eq:utilconv2}
\end{align}
where Eq.~(\ref{eq:utilconv1}) follows since $p_\text{loss}^\mathcal{Q}(K)=\text{O}(K^{-\beta_Q})$ for some $\beta_Q \geq 2$ {and $U_D(\cdot)$ is an analytic function on the non-negative real line}, and Eq.~(\ref{eq:utilconv2}) comes from choosing $\delta^{(a)}=\beta_Q \bar \sigma_a^2\frac{\log{K}}{K}$. This completes the proof of Theorem~\ref{th:bufferbattery}.}


From conservation of data, the incoming data is transmitted completely except for the amount lost due to data buffer overflows. Thus,
\begin{align}
&\rho^+(\lambda+\delta^{(a)})+(\rho^--p_\text{empty}^\mathcal{Q}(K))(\lambda-\delta^{(a)})\notag\\
&\qquad\qquad\qquad\qquad\qquad\qquad=\lambda(1-p_\text{loss}^\mathcal{Q}(K)).\label{eq:flowthm3}
\end{align}
By an extension of Lemmas~\ref{lemma:exprate2} and~\ref{lemma:s}, it can be shown that $p_\text{loss}^\mathcal{Q}(K)=\Theta\left(K^{-\beta_Q}\right)$. We can simplify Eq.~(\ref{eq:flowthm3}) as,
\begin{align}
\rho^+\delta^{(a)}-\rho^-\delta^{(a)}&=\Theta\left(K^{-\beta_Q}\right).\label{eq:firstorderthm3}
\end{align}
By substituting Eq.~(\ref{eq:firstorderthm3}) in the first-order term of Eq.~(\ref{eq:genutilthm3}), we observe that the scheme $\mathcal{Q}$ achieves $U_D(\mu)-\bar{U}_D^\mathcal{B}=\Theta\left(\frac{(\log K)^2}{K^2}\right)$.} This completes the proof of Theorem~\ref{th:bufferbattery}.}

\section{Proof for Theorem~\ref{th:expexprate}}\label{app:expexprate}

The proof of this theorem is constructive. We use the energy management scheme $\mathcal{E}$, given in Eq.~\ref{eq:thm4_scheme}. With this scheme, the mean drifts for the battery state and the data queue state are given by $\delta^{(r)} = \nu \left( \mu -C^{-1}(\lambda)\right)$ and $\delta^{(a)}=C(\mu-\delta^{(r)})-\lambda$, respectively. Since the limiting regime for the load is $\lambda\uparrow C(\mu)$, $\delta^{(r)} \downarrow 0$ and hence $C(\mu-\delta^{(r)})-\lambda\downarrow 0$. As a result, both the battery and the data queue will operate in the {heavy traffic} limit and we can apply the diffusion limits on these processes to obtain the required probability decay exponents.

{In the diffusion limit~\cite{WhittLimits,AsmussenQueueBook}, the decay rate for the discharge probability, $p_\text{underflow}^\mathcal{E}(M)=\lim_{t\to\infty}\Prob{\mathbf{B}(t)<0}$, for this energy management scheme can be calculated as,
\begin{align}
\lim_{M \to \infty} \ \lim_{\lambda\uparrow C(\mu)}\frac{1}{M \delta^{(r)}} \log p_\text{underflow}^\mathcal{E}(M)&=-\frac{2}{\bar\sigma_r^2},
\end{align}
where the decay exponent is found via the direct application of Theorem~7.1 in Chapter 10 of~\cite{AsmussenQueueBook}.
With an identical approach, by applying the diffusion limit to the data buffer process and substituting $\delta^{(a)}=C(\mu-\delta^{(r)})-\lambda$, we can find the decay rate for the buffer overflow probability, $p_\text{overflow}^\mathcal{E}(K)=\lim_{t\to\infty}\Prob{\mathbf{Q}(t)>K}$ as,
\begin{align}
\lim_{K\to \infty} \ \lim_{\lambda\uparrow C(\mu)} \frac{1}{K\delta^{(a)}}\log p_\text{overflow}^\mathcal{E}(K)&=-\frac{2}{\bar\sigma_a^2}.\label{eq:policyE2}
\end{align}}
\section*{Acknowledgment}
We thank the associate editor, Professor Lachlan Andrew for his valuable feedback and guidance throughout the review process.
\bibliographystyle{IEEEtran}
\bibliography{research}
\begin{IEEEbiography}[{\includegraphics[width=1in,height=1.25in,clip,keepaspectratio]{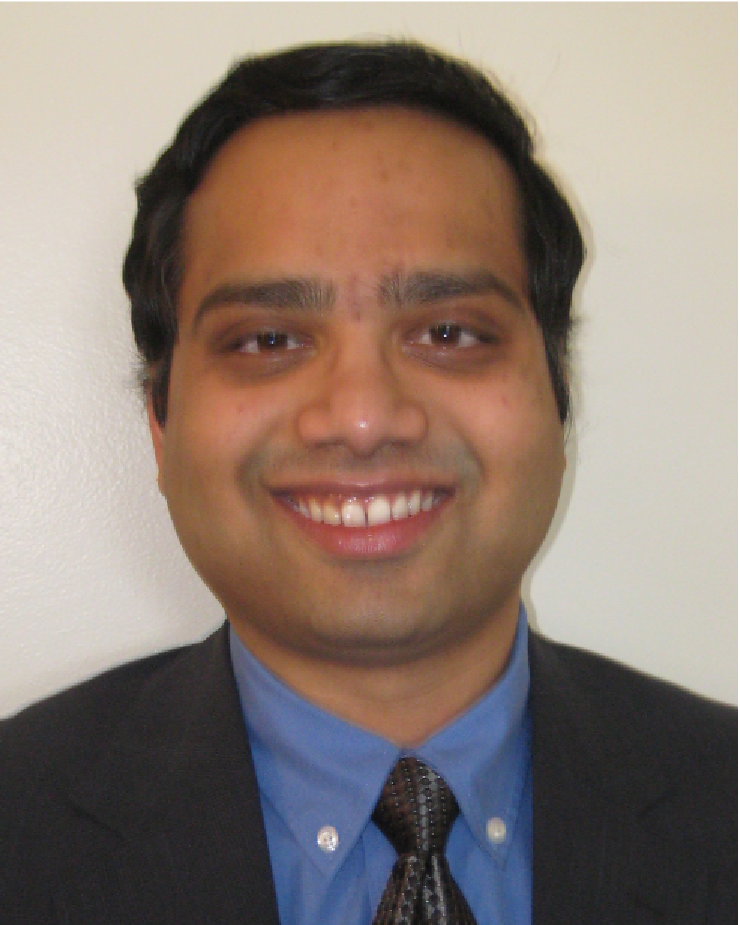}}]{Rahul Srivastava}
received the B.Tech. degree
in electrical engineering from the Indian Institute
of Technology, Madras, India, the M.S.
degree in electrical and computer engineering from
Rice University, Houston, USA, and the Ph.D. degree in electrical and computer engineering
from the Ohio State University, Columbus, USA in 2002, 2005, and 2010
respectively. He is currently working as a Staff Scientist in the Wireless Connectivity Group at the Broadcom Corporation, Sunnyvale, USA. His research interests include wireless communications, communication networks, stochastic processes, and optimization theory.
\end{IEEEbiography}
\begin{IEEEbiography}[{\includegraphics[width=1in,height=1.25in,clip,keepaspectratio]{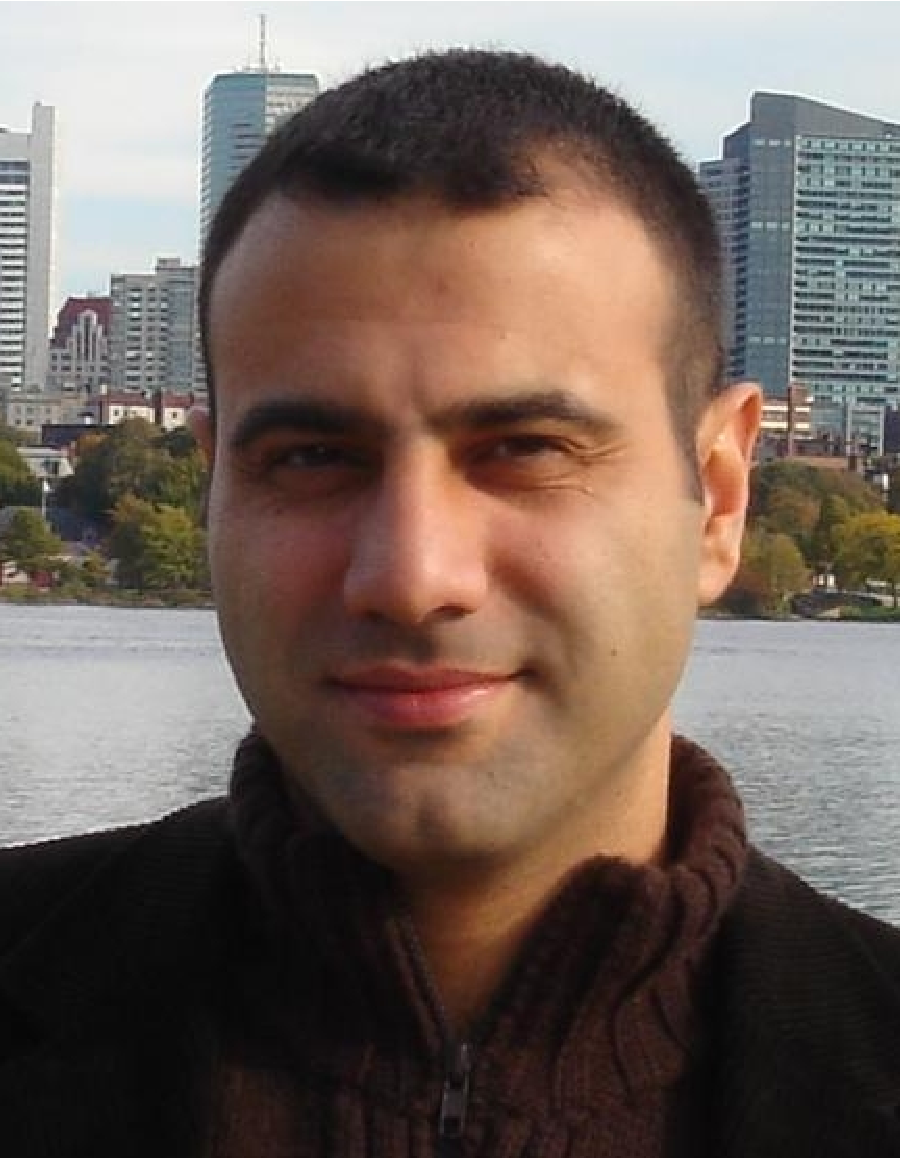}}]{Can Emre Koksal}
received the B.S. degree in electrical engineering from the Middle East Technical University, Ankara, Turkey, in 1996, and the S.M. and Ph.D. degrees from the Massachusetts Institute of Technology (MIT), Cambridge, in 1998 and 2002, respectively, in electrical engineering and computer science. He was a Postdoctoral Fellow in the Networks and Mobile Systems Group in the Computer Science and Artificial Intelligence Laboratory, MIT, until 2003 and a Senior Researcher jointly in the Laboratory for Computer Communications and the Laboratory for Information Theory at EPFL, Lausanne, Switzerland, until 2006. Since then, he has been an Assistant Professor in the Electrical and Computer Engineering Department, Ohio State University, Columbus, Ohio. His general areas of interest are wireless communication, communication networks, information theory, stochastic processes, and financial economics.

He is the recipient of the National Science Foundation CAREER Award in 2011, the OSU College of Engineering Lumley Research Award in 2011, and the co-recipient of an HP Labs - Innovation Research Award in 2011. 
\end{IEEEbiography}

\end{document}